\documentclass[a4paper,10pt,twoside]{cpc-hepnp}

\usepackage{lineno}
\usepackage{multirow}
\usepackage{color}
\usepackage{url}

\usepackage{multicol}
\usepackage{graphicx}
\usepackage{booktabs}
\usepackage{amssymb,bm,mathrsfs,bbm,amscd}
\usepackage[tbtags]{amsmath}
\usepackage{lastpage}
\usepackage{hyperref}

\usepackage[figuresright]{rotating}
\usepackage{longtable}
\usepackage{lineno}
\usepackage{hyperref}

\newcommand{\tabincell}[2]{\begin{tabular}{@{}#1@{}}#2\end{tabular}}
\usepackage{graphicx}  
\usepackage{subfigure}
\usepackage{epstopdf}
\usepackage{url}
\usepackage{multirow}
\usepackage{threeparttable}
\usepackage{lineno}

\graphicspath{{figures/}}

\begin{document}

\fancyhead[c]{\small Chinese Physics C~~~Vol. XX, No. X (202X)
XXXXXX} \fancyfoot[C]{\small XXXXXX-\thepage}


\title{Search For Electron-Antineutrinos Associated With Gravitational-Wave Events GW150914, GW151012, GW151226, GW170104, GW170608, GW170814, and GW170817 at Daya Bay}


\newcommand{\ECUST}{1}
\newcommand{\Wisconsin}{2}
\newcommand{\Yale}{3}
\newcommand{\BNL}{4}
\newcommand{\NTU}{5}
\newcommand{\IHEP}{6}
\newcommand{\NUU}{7}
\newcommand{\TsingHua}{8}
\newcommand{\SZU}{9}
\newcommand{\ZSU}{10}
\newcommand{\NCEPU}{11}
\newcommand{\CUHK}{12}
\newcommand{\Siena}{13}
\newcommand{\UCI}{14}
\newcommand{\USTC}{15}
\newcommand{\Charles}{16}
\newcommand{\UIUC}{17}
\newcommand{\LBNL}{18}
\newcommand{\IIT}{19}
\newcommand{\Dubna}{20}
\newcommand{\BNU}{21}
\newcommand{\XJTU}{22}
\newcommand{\UH}{23}
\newcommand{\CIAE}{24}
\newcommand{\SDU}{25}
\newcommand{\Guangxi}{26}
\newcommand{\VirginiaTech}{27}
\newcommand{\NCTU}{28}
\newcommand{\UC}{29}
\newcommand{\TempleUniversity}{30}
\newcommand{\DGUT}{31}
\newcommand{\UCB}{32}
\newcommand{\HKU}{33}
\newcommand{\NanKai}{34}
\newcommand{\SJTU}{35}
\newcommand{\Princeton}{36}
\newcommand{\CalTech}{37}
\newcommand{\WM}{38}
\newcommand{\NJU}{39}
\newcommand{\CGNPG}{40}
\newcommand{\NUDT}{41}
\newcommand{\IowaState}{42}
\newcommand{\CQU}{43}
\author{
F.~P.~An$^{\ECUST}$ \and 
A.~B.~Balantekin$^{\Wisconsin}$ \and
H.~R.~Band$^{\Yale}$ \and
M.~Bishai$^{\BNL}$ \and
S.~Blyth$^{\NTU}$ \and
G.~F.~Cao$^{\IHEP}$ \and 
J.~Cao$^{\IHEP}$ \and 
J.~F.~Chang$^{\IHEP}$ \and 
Y.~Chang$^{\NUU}$ \and 
H.~S.~Chen$^{\IHEP}$
% ...
\and 
}
\maketitle 
\address{
\vspace{0.3cm}
{\normalsize (Daya Bay Collaboration)} \\
\vspace{0.3cm}
$^{\ECUST}$Institute of Modern Physics, East China University of Science and Technology, Shanghai \\
$^{\Wisconsin}$University~of~Wisconsin, Madison, Wisconsin 53706 \\
$^{\Yale}$Wright~Laboratory and Department~of~Physics, Yale~University, New~Haven, Connecticut 06520 \\
$^{\BNL}$Brookhaven~National~Laboratory, Upton, New York 11973 \\
$^{\NTU}$Department of Physics, National~Taiwan~University, Taipei \\
$^{\IHEP}$Institute~of~High~Energy~Physics, Beijing \\
$^{\NUU}$National~United~University, Miao-Li \\
$^{\TsingHua}$Department~of~Engineering~Physics, Tsinghua~University, Beijing \\
$^{\SZU}$Shenzhen~University, Shenzhen \\
$^{\ZSU}$Sun Yat-Sen (Zhongshan) University, Guangzhou \\
$^{\NCEPU}$North~China~Electric~Power~University, Beijing \\
$^{\CUHK}$Chinese~University~of~Hong~Kong, Hong~Kong \\
$^{\Siena}$Siena~College, Loudonville, New York  12211 \\
$^{\UCI}$Department of Physics and Astronomy, University of California, Irvine, California 92697 \\
$^{\USTC}$University~of~Science~and~Technology~of~China, Hefei \\
$^{\Charles}$Charles~University, Faculty~of~Mathematics~and~Physics, Prague \\
$^{\UIUC}$Department of Physics, University~of~Illinois~at~Urbana-Champaign, Urbana, Illinois 61801 \\
$^{\LBNL}$Lawrence~Berkeley~National~Laboratory, Berkeley, California 94720 \\
$^{\IIT}$Department of Physics, Illinois~Institute~of~Technology, Chicago, Illinois  60616 \\
$^{\Dubna}$Joint~Institute~for~Nuclear~Research, Dubna, Moscow~Region \\
$^{\BNU}$Beijing~Normal~University, Beijing \\
$^{\XJTU}$Department of Nuclear Science and Technology, School of Energy and Power Engineering, Xi'an Jiaotong University, Xi'an \\
$^{\UH}$Department of Physics, University~of~Houston, Houston, Texas  77204 \\
$^{\CIAE}$China~Institute~of~Atomic~Energy, Beijing \\
$^{\SDU}$Shandong~University, Jinan \\
$^{\Guangxi}$Guangxi University, Nanning \\
$^{\VirginiaTech}$Center for Neutrino Physics, Virginia~Tech, Blacksburg, Virginia  24061 \\
$^{\NCTU}$Institute~of~Physics, National~Chiao-Tung~University, Hsinchu \\
$^{\UC}$Department of Physics, University~of~Cincinnati, Cincinnati, Ohio 45221 \\
$^{\TempleUniversity}$Department~of~Physics, College~of~Science~and~Technology, Temple~University, Philadelphia, Pennsylvania  19122 \\
$^{\DGUT}$Dongguan~University~of~Technology, Dongguan \\
$^{\UCB}$Department of Physics, University~of~California, Berkeley, California  94720 \\
$^{\HKU}$Department of Physics, The~University~of~Hong~Kong, Pokfulam, Hong~Kong \\
$^{\NanKai}$School of Physics, Nankai~University, Tianjin \\
$^{\SJTU}$Department of Physics and Astronomy, Shanghai Jiao Tong University, Shanghai Laboratory for Particle Physics and Cosmology, Shanghai \\
$^{\Princeton}$Joseph Henry Laboratories, Princeton~University, Princeton, New~Jersey 08544 \\
$^{\CalTech}$California~Institute~of~Technology, Pasadena, California 91125 \\
$^{\WM}$College~of~William~and~Mary, Williamsburg, Virginia  23187 \\
$^{\NJU}$Nanjing~University, Nanjing \\
$^{\CGNPG}$China General Nuclear Power Group, Shenzhen \\
$^{\NUDT}$College of Electronic Science and Engineering, National University of Defense Technology, Changsha \\
$^{\IowaState}$Iowa~State~University, Ames, Iowa  50011 \\
$^{\CQU}$Chongqing University, Chongqing \\
}


\begin{abstract}
Providing a possible connection between neutrino emission and gravitational-wave (GW) bursts is important to our understanding of the physical processes that occur when black holes or neutron stars merge. In the Daya Bay experiment, using the data collected from December 2011 to August 2017, a search has been performed for electron-antineutrino signals coinciding with detected GW events, including GW150914, GW151012, GW151226, GW170104, GW170608, GW170814, and GW170817. We used three time windows of $\mathrm{\pm 10~s}$, $\mathrm{\pm 500~s}$, and $\mathrm{\pm 1000~s}$ relative to the occurrence of the GW events, and a neutrino energy range of 1.8 to 100~MeV to search for correlated neutrino candidates. The detected electron-antineutrino candidates are consistent with the expected background rates for all the three time windows. Assuming monochromatic spectra, we found upper limits (90\% confidence level) on electron-antineutrino fluence of $(1.13~-~2.44) \times 10^{11}~\rm{cm^{-2}}$ at 5~MeV to $8.0 \times 10^{7}~\rm{cm^{-2}}$ at 100~MeV for the three time windows. Under the assumption of a Fermi-Dirac spectrum, the upper limits were found to be $(5.4~-~7.0)\times 10^{9}~\rm{cm^{-2}}$ for the three time windows.
\end{abstract}



\maketitle

\section{Introduction}
\label{introduction}

    The direct observation of gravitational waves (GWs) provides an important probe for investigating the dynamical origin of high-energy cosmic transients~\cite{Abadie:2010cf}. In 1987, Super-Kamiokande~\cite{Hirata:1987hu}, IMB~\cite{Bionta:1987qt}, and Baksan~\cite{Alekseev:1987ej} neutrino experiments observed neutrino signals 7 hours before the optical observation of a type II core-collapse supernova in the Large Magellanic Cloud. After the first detection of GW~\cite{TheLIGOScientific:2016agk} at the Advanced Laser Interferometer Gravitational-wave Observatory (LIGO)~\cite{Harry:2010zz}, astronomy has moved into another new and exciting era of exploration. The detected GW170817~\cite{TheLIGOScientific:2017qsa} was temporally correlated with the GRB170817A~\cite{GBM:2017lvd}, detected by the Fermi-GBR 1.7~s after the proposed coalescence of two neutron stars (NSs), providing the first direct evidence of a link between NS mergers and gamma-ray bursts (GRBs). Searching for coincident signals and exploring the dynamics of astronomical sources through distinct channels may enable the discrimination of competing model descriptions, thus expanding our understanding of the universe~\cite{Abadie:2010cf, Smith:2012eu, Guetta:2019kfn}. Therefore, active efforts for joint detection of GW events coinciding with neutrino signals have received substantial attention.

    The merging of two black holes (BHs), two NSs or BH with NS with rapidly rotating cores are expected to become GW sources with an output detectable by ongoing experiments~\cite{Bartos:2012vd}. NS-NS or NS-BH mergers with an accretion disk or BH accretion disks, that produce a gamma-ray burst (GRB)~\cite{Eichler:1989ve, Woosley:1993wj}, can also drive a relatively large neutrino outflow~\cite{Sekiguchi:2011zd, Kyutoku:2017wnb, Caballero:2011dw, Liu:2017kga, Caballero:2009ww}. Neutrinos and GWs can travel almost unchanged from their sources to a detector, acting as invaluable messengers regarding their hidden region of production.
    A list of GWs of interest for this study is listed in Table~\ref{tbl-1}, along with their observed times and distances.

\begin{table}[htbp]
\begin{center}
\caption{ GW events observed by the LIGO experiment~\cite{Abbott:2016blz, TheLIGOScientific:2016qqj, Abbott:2016nmj, Abbott:2017vtc, Abbott:2017gyy, Abbott:2017oio, TheLIGOScientific:2017qsa}.\label{tbl-1}}
\begin{tabular}{cccc}
\hline
\hline
GW events & Type of merged bodies& Detection time (UTC)& Distance $D_{\rm{LIGO}}$ (Mpc) \\
\hline
GW150914  & Black holes & 2015.09.14 09:50:45  & $410^{+160}_{-180}$  \\
GW151012  & Black holes & 2015.10.12 09:54:43  & $1100^{+500}_{-500}$ \\
GW151226  & Black holes & 2015.12.26 03:38:53  & $440^{+180}_{-190}$  \\
GW170104  & Black holes & 2017.01.04 10:11:58  & $880^{+450}_{-390}$  \\
GW170608  & Black holes & 2017.06.08 02:01:16  & $340^{+140}_{-140}$ \\
GW170814  & Black holes & 2017.08.14 10:30:43  & $540^{+130}_{-210}$  \\
GW170817  & Neutron stars  & 2017.08.17 12:41:04  &  $40^{+8}_{-14}$  \\
\hline
\hline
\end{tabular}
\end{center}
\end{table}


ANTARES and IceCube~\cite{Adrian-Martinez:2016xgn} searched for high-energy neutrino signals (above 100~GeV) within $\pm 500$~s relative to the occurrence of GW150914, but found no significant signal. KamLAND ~\cite{Gando:2016zhq} searched for low-energy neutrino signals (below 100~MeV) within $\pm 500$~s relative to GW150914 and GW151226 and also found no significant signal above background. Super-Kamiokande~\cite{Abe:2016jwn} searched for coincident neutrino events (3.5 MeV - 100 PeV) within $\pm 500$~s of GW150914 and GW151226, and observed no significant candidates beyond the expected background rate. Borexino~\cite{Agostini:2017pfa} searched for correlated neutrino events with energies exceeding 250~keV for GW150914, GW151226, and GW170104, obtaining a result consistent with the expected number of solar neutrino and background events. Thus far, no experiment has observed any connection between GW events and neutrino signals.

The Daya Bay experiment has been stably operating since 24 December 2011 and is part of the Supernova Early Warning System~\cite{Wei:2015qga} providing long-term monitoring of astrophysical electron-antineutrino ($\bar\nu_e$) bursts.
In this paper, we present a detailed off-line study using the data collected from 24 December 2011 to 30 August 2017~\cite{Adey:2018zwh}, searching for neutrino candidates coinciding with GW150914, GW151012, GW151226, GW170104, GW170608, GW170814, and GW170817.

\section{Experiment and Neutrino Detection}
\label{anti-detector}

\begin{figure}[]
  \centering
   \includegraphics[width=0.5\textwidth]{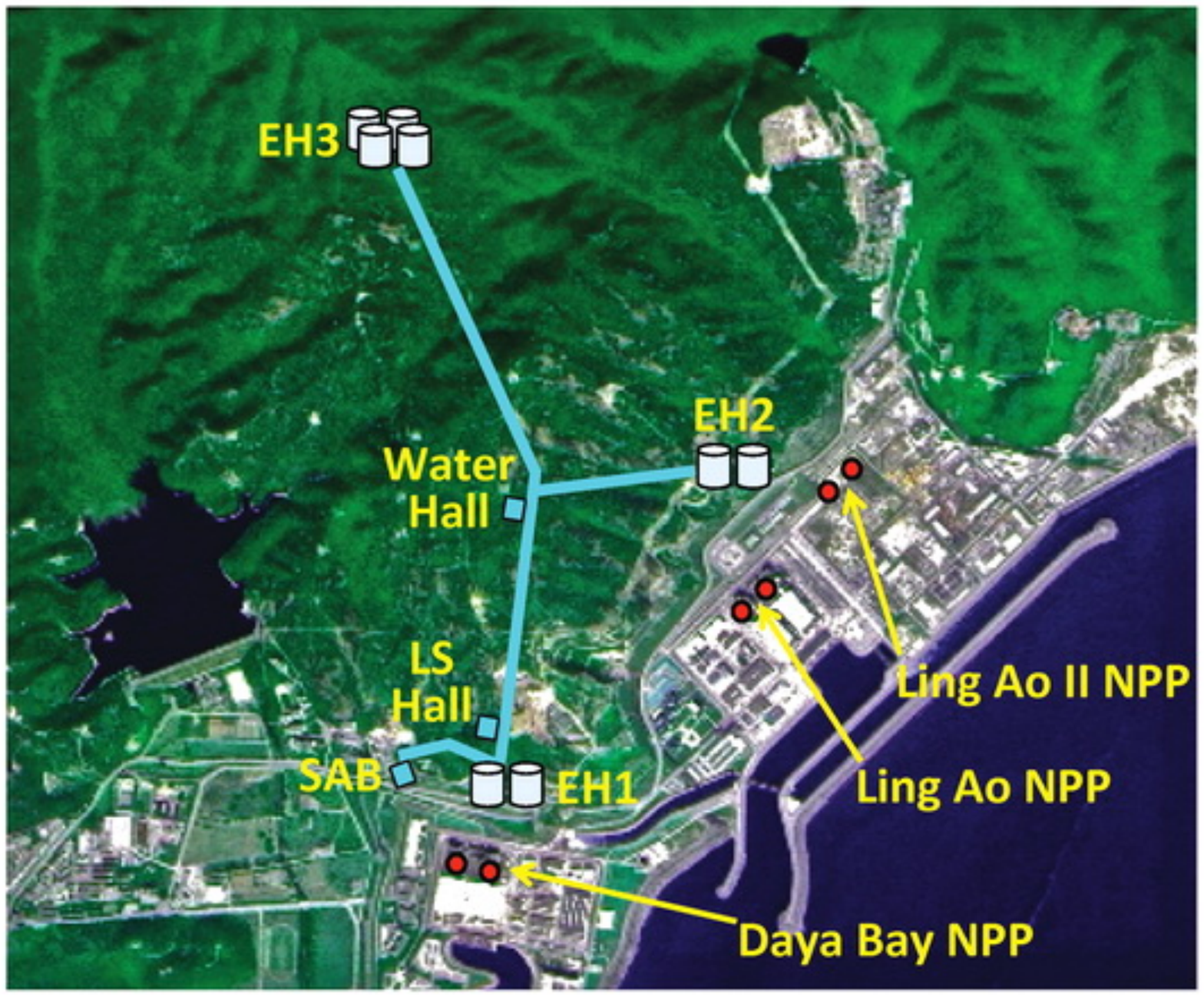}
  \caption{Layout of the Daya Bay experiment. Eight antineutrino detectors (ADs) are placed in three experimental halls (EHs). The solid cylinders represent the ADs. There are three nuclear power plants (NPPs): Daya Bay, Ling Ao, and Ling Ao II near EH1 and EH2. Each NPP consists of two reactor cores as the red dots in the figure.}\label{fig:EH}
\end{figure}

The Daya Bay Reactor Neutrino Experiment is designed to measure the electron-antineutrino disappearance with $\bar\nu_e$'s emitted by the six Daya Bay reactor cores. There are two nearby experimental halls (EH1 and EH2); each has two antineutrino detectors (ADs). Their distances to the nearby reactors range from 350 m to 600 m. A more distant hall (EH3) has four ADs, and their distances to the reactors range from 1500 m to 2000 m. The layout of the Daya Bay experiment is shown in Figure~\ref{fig:EH}. The three EHs are located underground with 250, 265, and 860 meters water-equivalent overburden, respectively.

\begin{figure}[]
  \centering
  \includegraphics[width=0.35\textwidth]{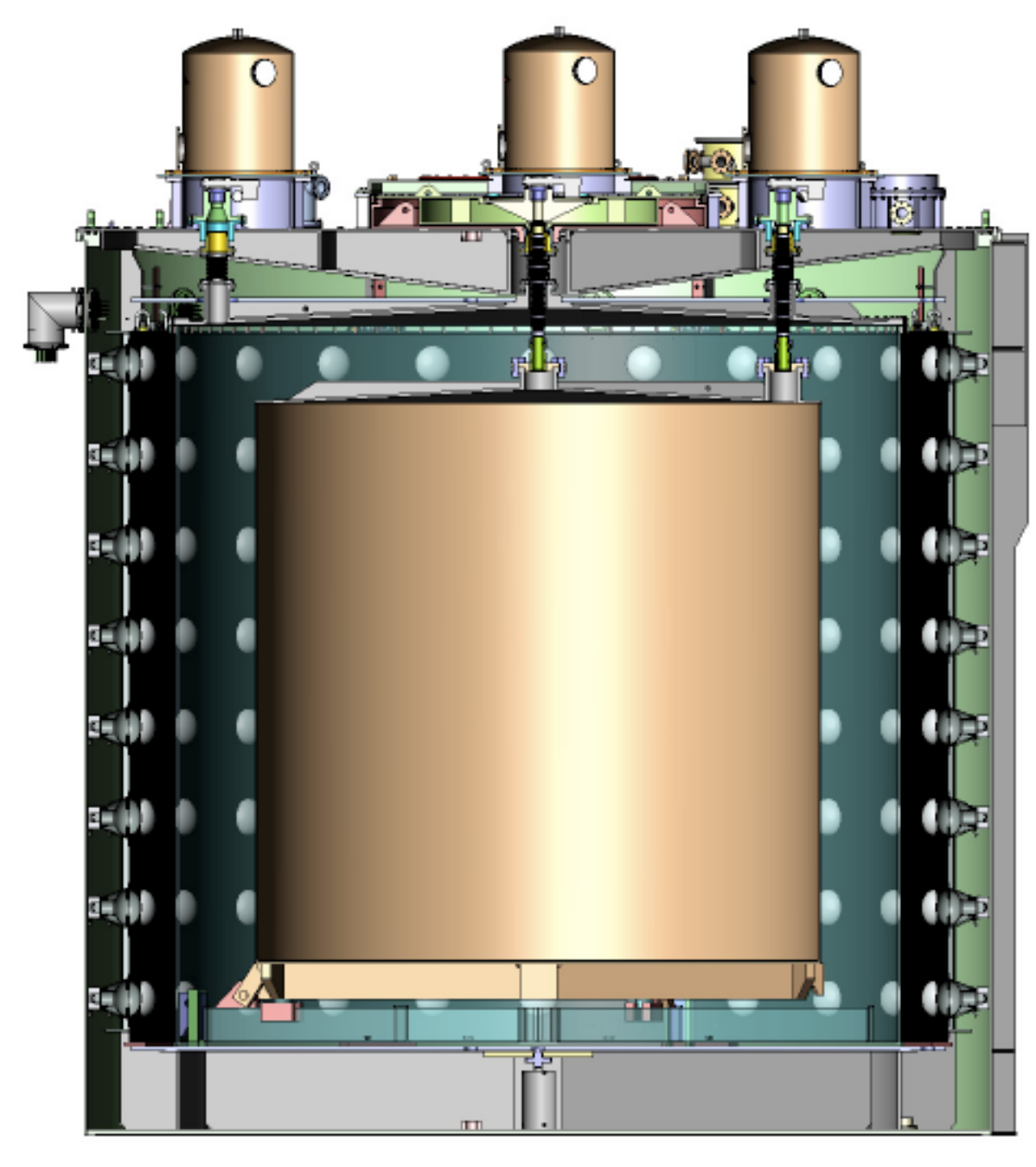}
  \caption{Schematic of a Daya Bay antineutrino detector. Three nested structures, starting
from the inner-most one, contain gadolinium-loaded liquid scintillator, un-doped liquid scintillator, and mineral oil. Three automated calibration units are installed at the top of each detector for calibration along three vertical axes. The inner surface of the stainless-steel vessel is equipped with 192 Hamamatsu R5912 PMTs to collect scintillation light.}
  \label{fig:ad}
\end{figure}

Each AD is a 5-m-diameter and 5-m-tall stainless-steel vessel containing three volumes separated by two coaxial transparent acrylic vessels.
The inner vessel holds 20 t of gadolinium-loaded liquid scintillator (Gd-LS) enclosed by the outer vessel filled with 22 t of undoped liquid scintillator (LS). The outermost volume holds 40 t of mineral oil and 192 20-cm-diameter photomultiplier tubes (PMTs).  An engineering plot of the AD is shown in Figure~\ref{fig:ad}. A more detailed description of the apparatus is available in~\cite{DayaBay:2012aa, An:2015qga}. The energy deposition and position of particles in each AD are reconstructed based on the amount of light collected by the PMTs. The universal time (UTC) of each event is recorded with a global positioning system (GPS) receiver.

The ADs are immersed in large water pools with at least 2.5 m of water on each side. Each water pool is divided into inner and outer regions (IWS and OWS) and both are instrumented with PMTs. The water is used to shield the detectors from environmental radiation and to tag muons via Cherenkov radiation.

 The reaction for detecting $\bar\nu_e$ at Daya Bay is the inverse beta decay (IBD):
\begin{linenomath}
\begin{equation}
 \label{eq:IBD}
 \bar\nu_e + p \rightarrow e^+ + n.
\end{equation}
\end{linenomath}
The $e^+$ rapidly deposits its kinetic energy and annihilates with an electron into two 0.511-MeV $\gamma$'s. The neutron thermalizes and captures on a Gd or H nucleus (nGd or nH). The energy deposited by the $e^+$ and neutron recoils forms a prompt signal, while the gamma ray(s) from the deexcitation after neutron capture gives a delayed signal. The prompt-delayed coincidence greatly suppresses the background. The IBDs that are temporally coincident with the GW events are searched for at the Daya Bay.

In the analysis, we required each AD to be in standard physics data acquisition around the time of each GW. For GW150914, GW151226, and GW151012, the experiment was in regular operation. During GW170104, both ADs in EH1 were being calibrated with radioactive sources and were excluded from the analysis. AD1 in EH1 was offline during GW170608, GW170814, and GW170817.

\section{Neutrino Fluence Measurement Method}
\label{analysis-method}

    We limited our search for $\bar\nu_e$ with energies below 100 MeV, as motivated by the 1987A supernova neutrino observation~\cite{Hirata:1987hu, Bionta:1987qt, Alekseev:1987ej} and some theoretical models in which the neutrino energies are mostly below 100~MeV~\cite{Caballero:2011dw, McLaughlin:2006yy}. This energy range is also within the best detection range for the Daya Bay detectors.

    Theoretically, the arrival sequence of neutrinos and GWs along with the duration of the neutrino burst depend on the physical process and modeling. To accommodate the uncertainties, we adopted multiple time windows to search for neutrino bursts associated with the GW events. First, we used a narrow time window of $\pm$10 s for GWs generated in a physical process similar to that of core-collapse supernovae~\cite{Raffelt:2012kt}. Next we applied an intermediate time window of $\pm$500 s to cover a greater time difference between the GW event and the predicted neutrino emission~\cite{Baret:2011tk}. Finally, we tested a more conservative time window of $\pm$1000 s~\cite{Fukuda:2002nf}.

    In the analysis, we first measured the electron-antineutrino fluence, $\Phi_{\rm{FD}}$, with a normalized, pinched Fermi-Dirac spectrum~\cite{Pinch1, Pinch2} with zero chemical potential and a pinching factor of $\eta =0$, as applied in the KamLAND experiment~\cite{Gando:2016zhq}. Using the number of electron-antineutrino candidates $N_{\nu}$ within the searching window, the electron-antineutrino fluence is calculated as
\begin{linenomath}
\begin{equation}
 \label{eq:FD}
 \Phi_{\rm{FD}} = \frac{N_{\nu}}{N_p\int \sigma (E_{\nu}) \epsilon(E_{\nu}) \phi(E_{\nu}) d E_{\nu}},
\end{equation}
\end{linenomath}
where $N_p$ is the number of target protons, $\sigma(E_\nu)$ is the IBD cross-section and $\epsilon(E_{\nu})$ is the detector efficiency.
The Fermi-Dirac spectrum is
\begin{linenomath}
\begin{equation}
\phi(E_{\nu})= \frac{1}{T^{3}F_{2}(\eta)}\frac{E_\nu^2}{e^{E_\nu/T-\eta}+1},
\end{equation}
\end{linenomath}
where the complete Fermi-Dirac integral function, $F_n(\eta)$, is given by
\begin{linenomath}
\begin{equation}
F_{n}(\eta)= \int^{\infty}_{0} \frac{x^{n}}{e^{x-\eta}+1}dx,
\end{equation}
\end{linenomath}
and the average $\bar{\nu}_e$ energy is set to $\langle E_{\nu}\rangle = 12.7$~MeV~\cite{Caballero:2015cpa} and $T=\langle E_{\nu}\rangle$/3.15.

To see the detailed sensitivity of the Daya Bay, the fluence was also estimated at several discrete energies $\Phi_{\rm{D}}(E_\nu)$ below 100 MeV as,
\begin{linenomath}
\begin{equation}
 \label{eq:delta}
 \Phi_{\rm{D}}(E_\nu) = \frac{N_{\nu}}{N_p\sigma (E_{\nu}) \epsilon(E_{\nu})},
\end{equation}
\end{linenomath}
where $N_{\nu}$ is the number of neutrino candidates as in Eq.~(\ref{eq:FD}), but limited to only the nearby region of the interested energy.

\section{Data Analysis}
\subsection{Energy conversion}\label{sec:energy}

In the IBD process, the positron kinetic energy depends on the true neutrino energy, $E_\nu$, and the scattering angle. In another word, the recoiling neutron from the IBD process shares part of the neutrino energy. Most of the neutron kinetic energy will transfer to the proton by scattering. The kinetic proton produces about half of the scintillation light given the same kinetic energy of an electron considering the quenching effect. The effects in the IBD process and the detector response are considered in the Daya Bay simulation and smearing of the reconstructed prompt energy, $E_{rec}$, is introduced when converted from $E_\nu$.

 In the analysis, for $E_\nu <100$ MeV, the relationship between the mean of the $E_{rec}$ (defined as $E_p$) and $E_\nu$ is estimated by simulation and is described by the following empirical equation:
\begin{linenomath}
\begin{equation}
E_p = A \times E^2_{\nu} + B \times E_\nu +C,
\label{Eq:Ep}
\end{equation}
\end{linenomath}
where $A= -0.0010~\rm{MeV^{-1}}$, $B= 1.01$, and $C= -0.73$ MeV. Due to the smearing effect in the energy conversion, various energy window will be opened to search neutrinos with specified discrete energies as described in next section.

The absolute energy scale given in Eq.~(\ref{Eq:Ep}) is calibrated by spallation neutrons, radiative sources, and Michel electrons.
At low energy, $E_p<10$ MeV, and in the Gd-LS region, the absolute energy scale uncertainty is less than 1\%~\cite{An:2016ses}, and in the LS region, the absolute energy scale uncertainty is better than 6\%~\cite{An:2016bvr}. At the higher end of Michel electrons, 53 MeV, the absolute energy scale uncertainty is better than 10\% for the whole AD.

\subsection{Candidate selection}\label{sec:selection}

    The IBD candidates are selected according to their features. Neutron capture on gadolinium emits an approximately 8-MeV $\gamma$ cascade, and the average neutron capture time in the Gd-LS region is 28~$\mu s$~\cite{An:2016ses}. A 2.2-MeV gamma is emitted after the neutron capture on hydrogen and the average capture time is 216~$\mu s$ in the LS region~\cite{An:2014ehw, An:2016bvr}.

    In the analysis, the standard selection criteria for IBDs in~\cite{An:2016bvr, An:2016ses} were adopted with a few minor modifications, as shown in Table~\ref{tb:cuts}. AD-triggered events caused by spontaneous light emission from PMTs (flashers) were rejected with no loss in efficiency~\cite{An:2012bu}. The coincidence time of the prompt and delayed signals was required to be greater than 1 $\rm{\mu s}$ and less than 200 and 400 $\rm{\mu s}$ for the nGd and nH samples, respectively. The delayed signal was required to be higher than 6 MeV for the nGd sample, and with a three-standard-deviation cut around the 2.2 MeV gamma-ray energy peak for the nH sample. Two more cuts were applied to suppress the accidental background in the nH sample.
    The distance between the prompt and delayed signal vertices was required to be less than 100 cm and a lower bound of 3.5 MeV was required for the reconstructed prompt energy.
    Furthermore the prompt-energy cut was adjusted according to the different searching regions as follows.

\begin{itemize}
\item For the Fermi-Dirac spectrum, the neutrino energy range of interest is 1.8 MeV to 100 MeV, which corresponds to the reconstructed prompt energy 0.7 MeV to 90 MeV as given by Eq.~(\ref{Eq:Ep}). Due to different ratios of signal to background, the searching energy range is further divided into two regions, low E ($E_{rec}$ $<$ 10 MeV) and high E ($E_{rec}$ $>$ 10 MeV).

\item For the monochromatic spectra, we selected $\bar{\nu}_e$ energies at 5, 7, 10, 20, 30, 50, 70, and 90~MeV to represent the entire energy range ($E_{\nu}<$ 100 MeV). For these specified discrete $\bar{\nu}_e$ energies, the prompt-energy search range was $E_p \pm \Delta$, where $\Delta$ defined as follows:
\begin{linenomath}
\begin{equation}
\Delta =  5 \times \sqrt{a^2 \times E^2_{p}+ {b^2} \times {E_{p}}+{c^2}},
\label{Eq:resolution}
\end{equation}
\end{linenomath}
where the parameters, $a=0.016$, $b=0.081$~$\rm{MeV^{1/2}}$, and $c=0.026$~MeV. We simply took the parameters from detector resolution equation~\cite{An:2016ses} to define signal window.

\end{itemize}

\begin{table*}[]
\caption{Selection criteria for the nH and nGd neutrino candidate samples. See the text for more details.}
\centering
\begin{tabular}{c|cc}
\hline
\hline
      & nGd & nH \\
\hline
Basic               &     \multicolumn{2}{c}{AD Trigger and flasher cut} \\
AD muon             &     \multicolumn{2}{c}{$>$ 100 MeV} \\
AD muon veto        &     \multicolumn{2}{c}{(0,~800) $\rm{\mu s}$} \\
Pool muon [IWS, OWS] &    \multicolumn{2}{c}{ $N_{\rm{IWS~PMT}}>12$ or $N_{\rm{OWS~PMT}}>15$ } \\
Pool muon veto      &     \multicolumn{2}{c}{(0,~600) $\rm{\mu s}$} \\
Shower muon         &     \multicolumn{2}{c}{$>$ 2.5 GeV} \\
Shower muon veto    &     \multicolumn{2}{c}{(0,~1) $\rm{s}$} \\
Coincidence time    & (1, 200) $\rm{\mu s}$ & (1, 400) $\rm{\mu s}$ \\
Delayed energy        &  $(6,~12)$ MeV  &  Peak $\pm ~ 3 \sigma_{E}$    \\
Coincidence distance &   N/A  & $<100$ cm   \\
Prompt energy (basic)&   N/A  & $>$ 3.5 MeV \\
Prompt energy (window)& \multicolumn{2}{c}{Signal searching region} \\
\hline
\hline
\end{tabular}
\label{tb:cuts}
\end{table*}

 The above selection criteria define different energy regions, neutron capture samples (nH or nGd), and ADs.
In total there are 32 data sets for the Fermi-Dirac spectrum study and 16 data sets for each monochromatic energy study.

 The number of candidates is measured within these regions of interest. The detailed time and energy information is documented in the Appendix~\ref{Appendix}.
One detailed example of GW150914 is shown in Table~\ref{tab:comparison}.

\begin{table}
  \centering
  \caption{Candidates and background details for the GW150914 coincidence neutrino search. Listed are the number of candidates in $\pm$500 s around GW150914. The number of background is calculated with data in $\pm$5 days or with the average background rate of all data multiplied by 1000 s. The uncertainties are statistical only.}\label{tab:comparison}
  \begin{tabular}{cccccc}
    \hline
    \hline
     &  &  nGd Low E &  nGd High E & nH Low E & nH High E   \\
  \hline

 \multirow{3}{*}{EH1-AD1}  & Candidate &  4   &  0  &  4  &  0   \\

                           & BKG. ($\pm5$ days) &   6.96 $\pm$ 0.08  &  0.060 $\pm$ 0.008  &   2.52 $\pm$ 0.06 &   0.080 $\pm$ 0.009  \\

                           & BKG. (Averaged) & 7.65 $\pm$ 0.01 & 0.064 $\pm$ 0.001 &  2.88 $\pm$ 0.01  &  0.092 $\pm$ 0.001 \\

\cline{2-6}
 \multirow{3}{*}{EH1-AD2}  & Candidate  & 5   &  0   &  1  &  0   \\

                           & BKG. ($\pm5$ days) &  6.95 $\pm$ 0.08  &  0.054 $\pm$ 0.007  &   2.54 $\pm$ 0.05 &  0.072 $\pm$ 0.008  \\
                           & BKG. (Averaged) & 7.65 $\pm$ 0.01 & 0.064 $\pm$ 0.001 &  2.88 $\pm$ 0.01  &  0.092 $\pm$ 0.001 \\

\cline{2-6}
 \multirow{3}{*}{EH2-AD1}  & Candidate  & 4  &  0   &  2   &  0    \\

                           & BKG. ($\pm5$ days) &   6.62 $\pm$ 0.08  & 0.037 $\pm$ 0.006  &   2.37 $\pm$ 0.05  &  0.041 $\pm$ 0.006     \\
                           & BKG. (Averaged) &    6.82 $\pm$ 0.01   &    0.043 $\pm$ 0.001  &   2.58 $\pm$ 0.01   &   0.063 $\pm$ 0.001  \\

 \cline{2-6}
 \multirow{3}{*}{EH2-AD2}  & Candidate  & 8  &  0   &  1   &  0    \\

                           & BKG. ($\pm5$ days) &   6.46 $\pm$ 0.08  &   0.027 $\pm$ 0.005  &   2.35 $\pm$ 0.05  &   0.056 $\pm$ 0.006  \\
                           & BKG. (Averaged) &      6.82 $\pm$ 0.01   &    0.043 $\pm$ 0.001  &   2.58 $\pm$ 0.01   &   0.063 $\pm$ 0.001  \\

 \cline{2-6}
 \multirow{3}{*}{EH3-AD1}  & Candidate  & 0  &  0   &  0   &  0    \\

                           & BKG. ($\pm5$ days) &   0.97 $\pm$ 0.03  &   0.004 $\pm$ 0.002  &   0.37 $\pm$ 0.02 &   0.008 $\pm$ 0.003  \\
                           & BKG. (Averaged) &     0.850 $\pm$ 0.001  &    0.0038 $\pm$ 0.0001 &    0.330 $\pm$ 0.001  &   0.0056 $\pm$ 0.0001  \\

  \cline{2-6}
 \multirow{3}{*}{EH3-AD2}  & Candidate  & 0  &  0   &  0   &  0    \\

                           & BKG. ($\pm5$ days) &   1.00 $\pm$ 0.03  &   0.003 $\pm$ 0.002  &   0.36 $\pm$ 0.02  &  0.007 $\pm$ 0.003  \\
                           & BKG. (Averaged) &     0.850 $\pm$ 0.001  &    0.0038 $\pm$ 0.0001 &    0.330 $\pm$ 0.001  &   0.0056 $\pm$ 0.0001  \\

    \cline{2-6}
 \multirow{3}{*}{EH3-AD3}  & Candidate  & 0  &  0   &  0   &  0    \\

                           & BKG. ($\pm5$ days) &  0.97 $\pm$ 0.03  &   0.001 $\pm$ 0.001  &   0.34 $\pm$ 0.02 &   0.004 $\pm$ 0.002  \\
                           & BKG. (Averaged) &     0.850 $\pm$ 0.001  &    0.0038 $\pm$ 0.0001 &    0.330 $\pm$ 0.001  &   0.0056 $\pm$ 0.0001  \\

   \cline{2-6}
 \multirow{3}{*}{EH3-AD4}  & Candidate  & 1  &  0   &  0   &  0    \\

                           & BKG. ($\pm5$ days) &  0.97 $\pm$ 0.03 &   0.007 $\pm$ 0.002  &  0.36 $\pm$ 0.02  &  0.005 $\pm$ 0.002  \\
                           & BKG. (Averaged) &     0.850 $\pm$ 0.001  &    0.0038 $\pm$ 0.0001 &    0.330 $\pm$ 0.001  &   0.0056 $\pm$ 0.0001  \\

  \hline
  \hline
  \end{tabular}
\end{table}

\subsection{Background}

 This section describes the background situation and how it is determined.
In all possible cases, we used the realtime estimation, i.e.~the result in $\pm$5 days around a GW, for background estimation.
But when the background is very low, we used the average of the entire data set.

\begin{figure}[h]
\centering
\includegraphics[width=0.5 \textwidth]{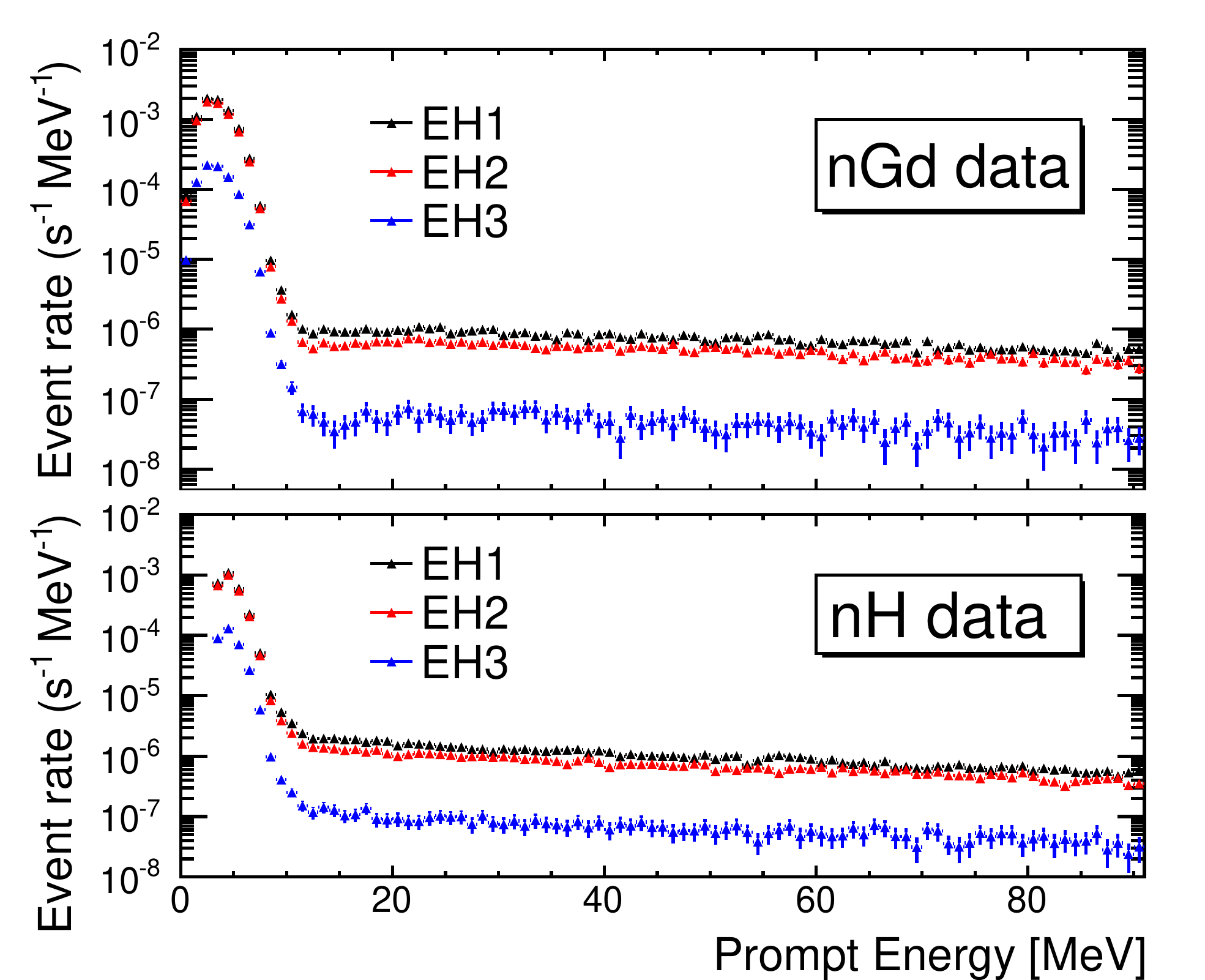}
\caption{Prompt-energy spectra of the neutrino candidates. The upper (lower) panel shows the spectra for the nGd (nH) selection.}
\label{Fig::Spectra}
\end{figure}

The prompt-energy spectra for the entire analyzed data period after the selection criteria are treated as background and shown in Figure~\ref{Fig::Spectra}. In the low-energy region ($<$10 MeV), the background is dominated by reactor anti-neutrinos because the detector is close to the reactors. Their variation with time follows the trend of the reactor power, which was published in a previous Daya Bay paper~\cite{An:2017osx}. In the high-energy region (10 to 100 MeV), the background is dominated by fast neutrons. The fast neutrons are the spallation products induced by cosmic-ray muons that are not vetoed. The proton recoils of a neutron introduce a prompt signal, and the capture of the neutron is the delayed signal~\cite{An:2016ses, An:2016bvr}. The muon flux has some seasonal changes and the maximal annual change is less than 1\%~\cite{An:2017wbm}; for the short time span of this study, the variation of these backgrounds is not significant.

The detailed background determined by the $\pm$5 days data is shown in Table~\ref{tab:comparison} for GW150914 as an example.
The average background rates for each AD are calculated with all data and shown in Table~\ref{tb:bkg}.
The ADs in the same EH are so close to each other and we will also compile other results in the same EH in the following for conciseness.
In Table~\ref{tab:comparison}, the background rate calculated by the average rate multiplied by 1000 s is also shown for comparison.

\begin{table}[]
    \centering
    \caption{ Average background rate (per second per antineutrino detector) for the studied energy spectrum (Low E: $E_{rec} < 10$ MeV, High E: $E_{rec}> 10$ MeV). The uncertainties are statistical only. }
    \label{tb:bkg}
    \begin{tabular}{cccc}
    \hline
    \hline
              &  EH1 &  EH2 & EH3   \\
  \hline
              & \multicolumn{3}{c}{nGd}  \\
        Low E   &  $(7.65 \pm 0.01)~\times 10^{-3}$ &  $(6.82 \pm 0.01)~\times 10^{-3}$ & $(8.45 \pm 0.01)~\times 10^{-4}$ \\
        High E  &  $(6.35 \pm 0.04 )~\times 10^{-5}$ &  $(4.32\pm  0.04)~\times 10^{-5}$ & $(3.83 \pm 0.08)~\times 10^{-6}$ \\

              &  \multicolumn{3}{c}{nH}  \\

         Low E   &  $(28.75 \pm 0.04)~\times 10^{-4}$ &  $(25.76 \pm 0.03)~\times 10^{-4}$ & $(3.25 \pm 0.01)~\times 10^{-4}$ \\
         High E  &  $(9.20 \pm  0.05 )~\times 10^{-5}$ &  $(6.30 \pm 0.05 )~\times 10^{-5}$ & $(5.65 \pm 0.10)~\times 10^{-6}$ \\
  \hline
  \hline
    \end{tabular}
\end{table}

Depending on the statistics (Figure~\ref{Fig::Spectra}, Table~\ref{tab:comparison}, and Table~\ref{tb:bkg}), for low E region, where the background rate is high and sensitive to reactor power changes, the background rate is determined with the data of $\pm$5 days of a GW. For the high E region, the background rate is low, sometimes there is zero background event in $\pm$5 days for the EH3, and considering the stableness of the muon flux, the background prediction is given by scaling the average background rate.

\subsection{Candidates and background comparison}
 
The following is a detailed comparison of the background and candidates.
For example, the number of candidates for GW150914 in $\pm$500 s for the 32 regions are shown in Table~\ref{tab:comparison}. The information can also be read off from the figures in the appendix~\ref{Appendix}.
The number of background comes from the result of $\pm5$-day calculation and the result by scaling the average rates in Table~\ref{tb:bkg} by 1000 s. The numbers of candidates agree with the statistical fluctuation of the background predictions.

 A thorough check shows that, for all GW candidates and all searching regions, the numbers of candidates agree with the background predictions within their statistical fluctuation. Given the current situation that there is no statistically significant signal found, the detailed background information is not listed.

\subsection{Detection efficiency}\label{ADE}

The signal detection efficiency, $\epsilon$, is defined as~\cite{An:2016ses, An:2016bvr}:
\begin{linenomath}
\begin{equation}
\label{eq:eff}
 \epsilon = \epsilon_\mu \cdot \epsilon_m \cdot \epsilon_{other},
\end{equation}
\end{linenomath}
with
\begin{linenomath}
\begin{equation}
 \epsilon_{other} = \sum_v (N_{p,v} \cdot \epsilon_{Ep,v} \cdot \epsilon_{Ed,v} \cdot \epsilon_{D,v} \cdot \epsilon_{T,v}) / \sum_v N_{p,v},
\end{equation}
\end{linenomath}
where $\epsilon_\mu$ is the muon veto efficiency, $\epsilon_m$ is the multiplicity cut efficiency for the two-fold event selection, and $\epsilon_{Ep,v}$, $\epsilon_{Ed,v}$, $\epsilon_{D,v}$, and $\epsilon_{T,v}$ correspond to the prompt energy, delayed energy, coincident distance (for the nH sample only), and coincident time efficiency, respectively. The efficiency $\epsilon_{other}$ is evaluated for each detector volume $v$ separately and is summed according to the number of free protons $N_{p,v}$ in each volume.

    The $\epsilon_\mu$ accounts for the live time lost due to the application of the muon veto time. We averaged $\epsilon_{\mu}$ for $\pm5$ days around each GW arrival time for the final result. The average values of $\epsilon_\mu$ over all GW candidates for the nGd and nH sample selection are shown in Table~\ref{tbl-efficiency}.

    The multiplicity cut efficiency $\epsilon_m$ applies to the two-fold event selection. The quantity $\epsilon_m$ is a function of the event rate and has a minor dependence on the muon rate and coincident time. Further details can be found in~\cite{Yu:2015pka}. In this study, we averaged the $\epsilon_m$ for $\pm5$ days around each GW arrival time. The differences between the nGd and nH samples and between the halls are not significant. The average values of $\epsilon_m$ over all GW candidates for the nGd and nH sample selection are shown in Table~\ref{tbl-efficiency}.

\begin{figure}[]
\flushleft
\centering
\includegraphics[width=0.5 \textwidth]{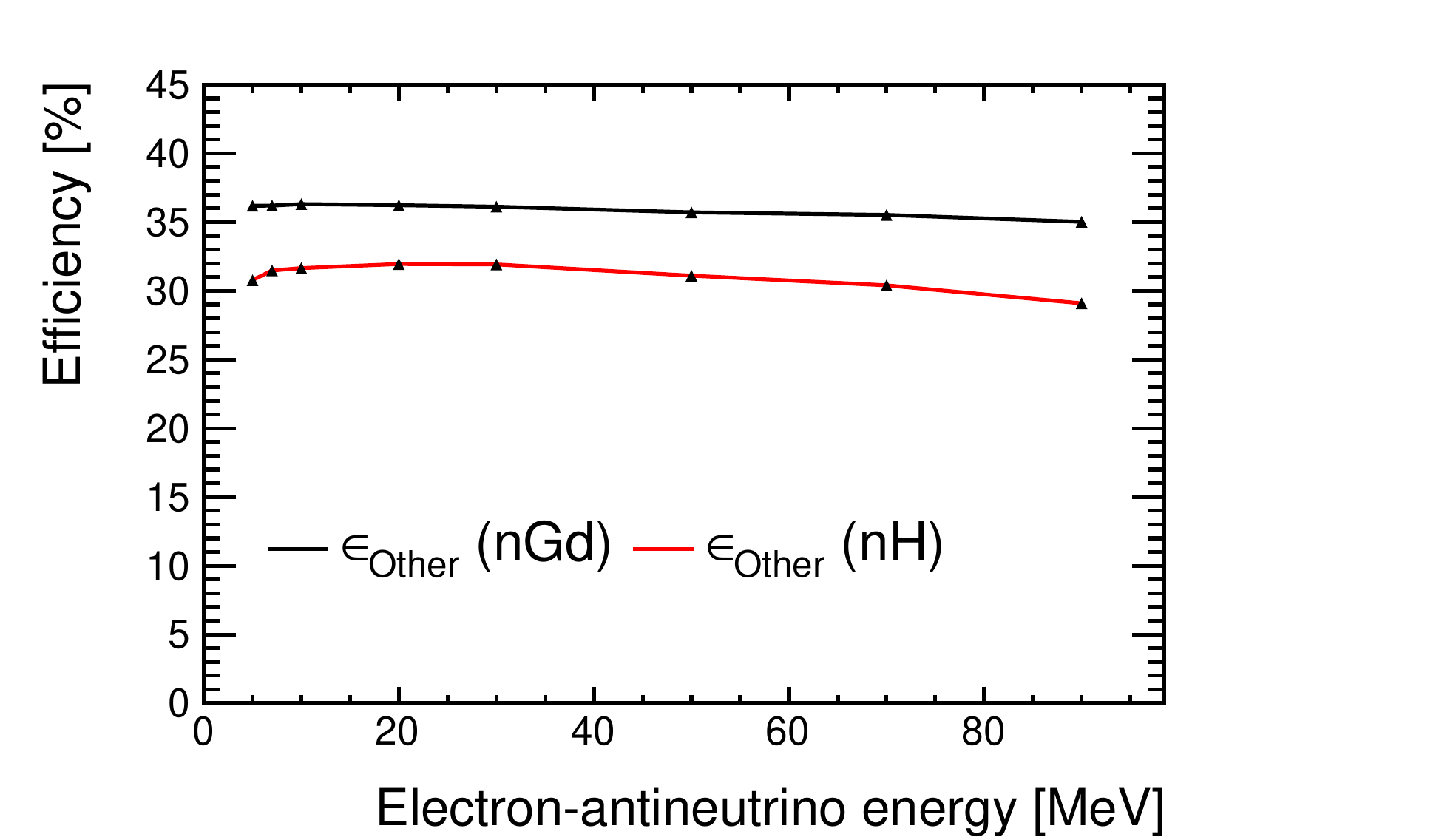}
\caption{IBD selection efficiency $\epsilon_{other}$ estimated with MC simulation as a function of the $\bar{\nu}_e$ energy.}
\label{Fig:eff_other}
\end{figure}

    For $\epsilon_{other}$, all values are estimated by simulation.
    The efficiencies $\epsilon_{other}$ for the Fermi-Dirac study are shown in Table~\ref{tbl-eother} and for monochromatic study in Figure~\ref{Fig:eff_other}. The prompt and delayed energy cuts for the Fermi-Dirac and monochromatic spectra leads to the main changes for $\epsilon_{other}$. At high energy, the efficiency decreases because of higher neutron energy and other neutron inelastic scattering. For the nH sample, at low energy, the efficiency decreases because of the 3.5 MeV prompt-energy cut.

    The uncertainty of the energy cut efficiency can be estimated by the absolute energy scale uncertainty (see section~\ref{sec:energy}) and is less than 4\%. The simulation of coincident distance and time between the prompt and delayed signal is validated with various natural coincident signals at low energy in~\cite{An:2016bvr} and with fast neutrons at high energy and the efficiency uncertainty is less than 10\%. The total uncertainty of efficiency is less than 10\% and negligible for calculating the upper limit.

\begin{table*}[]
    \caption{Muon veto cut efficiency, $\epsilon_\mu$, and multiplicity efficiency, $\epsilon_m$. \label{tbl-efficiency} }
    \centering
    \begin{tabular}{ccccc}    \hline    \hline
            &   &  EH1 &  EH2 & EH3   \\  \hline
      \multirow{2}{*}{nGd}                     &   $\epsilon_{\mu}$  &  79\% &   84\% & 98\%  \\
                                               &   $\epsilon_{m}$   &  98\% &   98\% &  98\% \\
    \multirow{2}{*}{nH}                      &    $\epsilon_{\mu}$  &   77\% &   81\% &  98\% \\
                                              & $\epsilon_{m}$   &   98\% &   99\% &  98\% \\  \hline  \hline
    \end{tabular}
\end{table*}

\begin{table}[]
    \centering
    \caption{ Summary of the $\epsilon_{other}$ for IBD searching with Fermi-Dirac spectrum for
     Low E and High E. \label{tbl-eother} }
    \begin{tabular}{ccc}\hline\hline
         & $E_p$$<$10 MeV   & $E_p$$>$10 MeV     \\\hline
     nGd & 4.3\% & 32\%  \\
     nH  & 4.4\% & 28\%  \\\hline \hline
    \end{tabular}
\end{table}

\section{Upper Limits on Electron-Antineutrino Fluence}\label{fluence}

 Given the observations in the study, the distributions of the GW $\bar{\nu}_e$ candidates in time and energy are consistent with the expected background. Thus, a maximum-likelihood fitting approach was used to calculate the upper limit according to the combination of candidates and backgrounds in the different searching regions.

\subsection{IBD cross-section} \label{cross-section}
   The IBD cross-section is taken from~\cite{Strumia:2003zx}. The average cross-section, $\bar{\sigma}$, is determined by integrating the IBD cross-section from 1.8 to 100 MeV over the neutrino spectrum. The detailed cross-section for the monochromatic energy and the average cross-section within 100~MeV for the Fermi-Dirac spectrum are given in Table~\ref{tbl-cross-section}.

 \begin{table*}[]
    \caption{IBD cross-section for the monochromatic spectra and the Fermi-Dirac spectrum.   \label{tbl-cross-section} }
    \centering
    \begin{tabular}{c|cccccccc}
    \hline
    \hline

  Cross-Section($\times 10^{-42} ~ \rm{cm^2}~$)& 5~MeV  & 7~MeV  & 10~MeV & 20~MeV & 30~MeV & 50~MeV & 70~MeV & 90~MeV \\
\hline
           $\sigma(E_\nu)$   & 1.27  & 2.96 & 6.76 & 28.9  & 63.0 & 156 & 268 & 389 \\
\hline
       $\bar{\sigma}$  &   \multicolumn{8}{c}{14.7 } \\

\hline
\hline
    \end{tabular}
\end{table*}

\subsection{Maximum-likelihood fit}
A Poisson probability can be calculated for each searching region $i$:
\begin{linenomath}
\begin{equation}
P_i(\Phi) = \frac{(N_i+b_i)^{n_i}}{n_{i}!} e^{-(N_i+b_i)},
\end{equation}
\end{linenomath}
where $N_i$ is the expected number of neutrino events within the searching region $i$ estimated for the expected fluence $\Phi$ (see Eq.~(\ref{eq:FD}) or Eq.~(\ref{eq:delta})), efficiency, cross-section and number of protons~\cite{An:2016ses, An:2016bvr},
$b_i$ is the expected background events, which is derived from the expected background rate. and $n_i$ is the number of observed neutrino candidates.
Because the dominant errors are statistical, no other errors of efficiency and background are included.
The combined likelihood of all data sets is
\begin{linenomath}
\begin{equation}
L  \left(\Phi \right)= \prod^{N_{\rm{ data ~ sets}}}_{i=1} P_i \left(\Phi\right).
\end{equation}
\end{linenomath}
Using the combined likelihood function, we constructed a test statistic based on the profile likelihood ratio, which can be used for a one-sided test for finding an upper limit as in Ref~\cite{Cowan:2010js}. We got the distribution of the test statistic using Monte-Carlo and finally deduced the upper limit of $\Phi(E_\nu)$.

\section{Results}\label{FAL}
\subsection{Fluence}
\begin{figure}[]
\flushleft
\centering
\includegraphics[width=0.5 \textwidth]{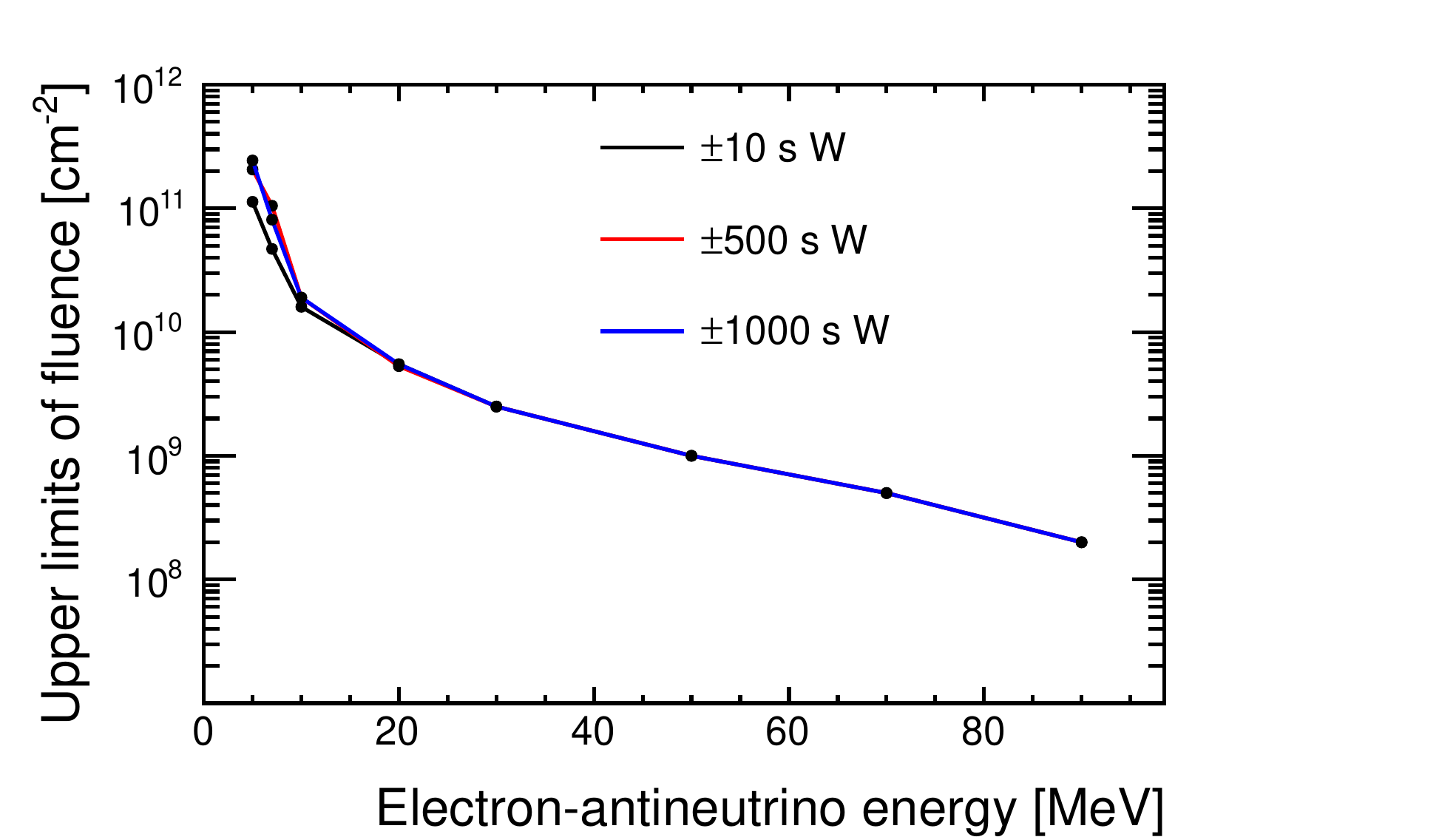}
\caption{ Upper limit (90\% C.L.) of the $\bar{\nu}_e$ fluence for each energy point for the GW-coincident event search. Three curves are shown for the three time windows (W).}
\label{Fig::Sen}
\end{figure}

    The upper limits (90\% C.L.) on the fluence with the Fermi-Dirac spectrum assumption for the searching time window of $\pm$500~s are shown in Table~\ref{tbl-500s} for each GW event. The remaining results are compiled in Table~\ref{tbl-upper} and Figure~\ref{Fig::Sen} which show typical $\bar{\nu}_e$ fluences estimated with the average numbers of candidates and background. For $E_\nu \geqslant 30$~MeV, the upper limits of the monochromatic spectrum fluence are identical for the three time windows, because the numbers of candidates and backgrounds are both close to null. The other variation are consistent with the expected efficiency changes and statistical fluctuations. Because multiple ADs are used at Daya Bay, the background within a single AD has almost no impact on the other ADs; hence a gain in sensitivity~\cite{Wei:2015coa}. The sensitivity is comparable to the KamLAND experiment~\cite{Gando:2016zhq}.

\begin{table}[]
\begin{center}
\renewcommand\arraystretch{1.5}
\caption{Upper limit (90\% C.L.) of the $\bar{\nu}_e$ fluence and luminosity, assuming a Fermi-Dirac spectrum within a search time window of $\pm 500$~s.\label{tbl-500s}}
\begin{tabular}{ccc}
\hline
\hline
  \tabincell{c}{ GW event} & \tabincell{c}{$\Phi_{\rm{FD}}$ $(\times 10^{10}~\rm{cm^{-2}})$ }  & \tabincell{c}{ $L_{\rm{GW}}$ ($\times 10^{60}$~erg) } \\
\hline
 GW150914    & 0.30   & 1.23  \\
 GW151012    & 0.79   & 23.3  \\
 GW151226    & 0.82   & 3.86  \\
 GW170104    & 0.97   & 18.3  \\
 GW170608    & 0.42   & 1.18  \\
 GW170814    & 0.73   & 5.18  \\
 GW170817    & 0.85   & 0.03  \\
\hline
\hline
\end{tabular}
\end{center}
\end{table}

 \begin{table*}[]
    \caption{Upper limits (90\% C.L.) for three search time windows. The results are estimated with the average numbers of candidates and backgrounds for the GW candidates.
    \label{tbl-upper}}
      {\footnotesize
    \begin{tabular}{c|cccccccc|c}
    \hline
    \hline
  Fluence ($\times 10^{10}~\rm{cm^{-2}}~$) & \multicolumn{8}{c|}{Monochromatic Spectra} & Fermi-Dirac Spectrum \\\hline
       $E_\nu$ &  5~MeV  & 7~MeV  & 10~MeV & 20~MeV & 30~MeV & 50~MeV & 70~MeV & 90~MeV &   (1.8, 100) MeV \\ \hline
   $\pm 10$~s  &  11.3   &  4.7    & 1.6   & 0.55   & 0.25   & 0.10   & 0.05   &0.02    &    0.70    \\
   $\pm 500$~s &  20.6   &  10.5   & 1.9   & 0.53   & 0.25   & 0.10   & 0.05   &0.02    &    0.63    \\
  $\pm 1000$~s &  24.4   &  8.1    & 1.9   & 0.55   & 0.25   & 0.10   & 0.05   &0.02    &    0.54    \\
  \hline
  \hline
    \end{tabular}
    }
\end{table*}

\subsection{Luminosity}
    The upper limits of the fluence can be converted into limits on total energy radiated in the form of $\bar{\nu}_e$'s for the mergers. With the assumption that it is the Fermi-Dirac spectrum in the energy range 1.8 - 100~MeV and the $\bar{\nu}_e$'s emanating from the source are isotropic, the total luminosity can then be expressed as
\begin{linenomath}
\begin{equation}
L_{\rm{GW}} = \Phi_{\rm{FD}} \times 4 \pi D_{\rm{LIGO}}^2 \times \langle E_{\nu}\rangle,
\end{equation}
\end{linenomath}
where $\langle E_{\nu}\rangle$ is the average $\bar{\nu}_e$ energy for the Fermi-Dirac spectrum (see Section~\ref{analysis-method}) and $D_{\rm{LIGO}} $ is the central value of the distance from the GW source to the Earth as given in Table~\ref{tbl-1}. The upper limits in luminosity $L_{\rm{GW}}$ for all of the GW events are listed in Table~\ref{tbl-500s}.

\section{Conclusion}\label{conclusion}

    We have searched for possible $\bar{\nu}_e$ signals with energies of 1.8 to 100~MeV coinciding with GW150914, GW151012, GW151226, GW170104, GW170608, GW170814, and GW170817 by assuming a Fermi-Dirac spectrum and monochromatic spectra. No candidate event above the background was found for any of the GW events in time windows reaching $\pm1000$~s. We used a maximum-likelihood fit to derive the 90\% C.L. upper limits for the $\bar{\nu}_e$ fluence, providing a comprehensive search across all of the observed GW events.

\section{Acknowledgements}
\label{Acknowledgements}
Daya Bay is supported in part by the Ministry of Science and
Technology of China, the U.S. Department of Energy, the Chinese
Academy of Sciences, the CAS Center for Excellence in Particle
Physics, the National Natural Science Foundation of China, the
Guangdong provincial government, the Shenzhen municipal government,
the China General Nuclear Power Group, Key Laboratory of Particle and
Radiation Imaging (Tsinghua University), the Ministry of Education,
Key Laboratory of Particle Physics and Particle Irradiation (Shandong
University), the Ministry of Education, Shanghai Laboratory for
Particle Physics and Cosmology, the Research Grants Council of the
Hong Kong Special Administrative Region of China, the University
Development Fund of The University of Hong Kong, the MOE program for
Research of Excellence at National Taiwan University, National
Chiao-Tung University, and NSC fund support from Taiwan, the
U.S. National Science Foundation, the Alfred~P.~Sloan Foundation, the
Ministry of Education, Youth, and Sports of the Czech Republic, the Charles University GAUK Proj. No. 284317,
the Joint Institute of Nuclear Research in Dubna, Russia, the National Commission of Scientific and
Technological Research of Chile, and the Tsinghua University
Initiative Scientific Research Program. We acknowledge Yellow River
Engineering Consulting Co., Ltd., and China Railway 15th Bureau Group
Co., Ltd., for building the underground laboratory. We are grateful
for the ongoing cooperation from the China General Nuclear Power Group
and China Light and Power Company.

\appendix
\section{Candidate distribution}
\label{Appendix}
\begin{figure}[htbp]
\centering
\includegraphics[width=0.27 \textwidth]{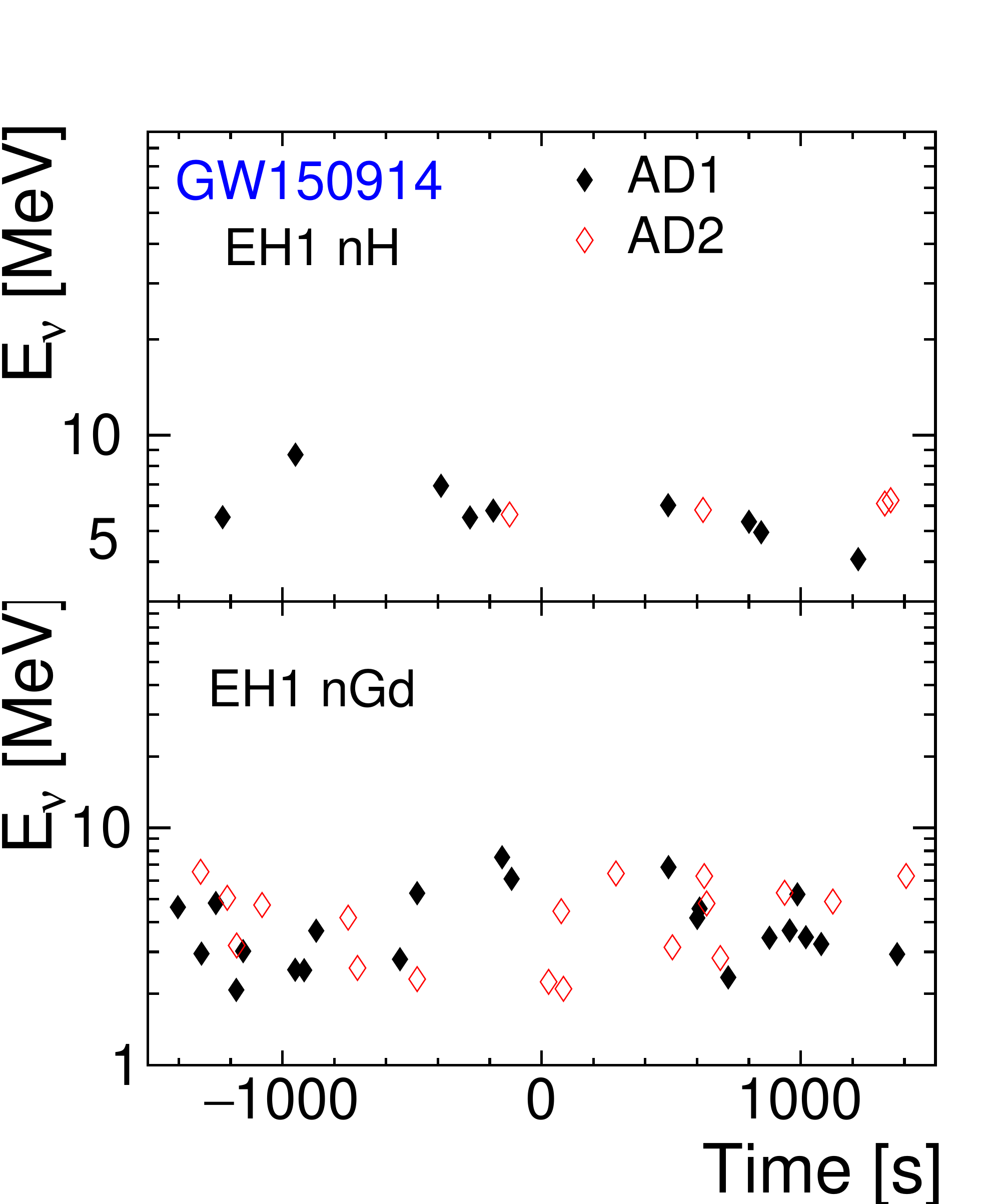}
\includegraphics[width=0.27 \textwidth]{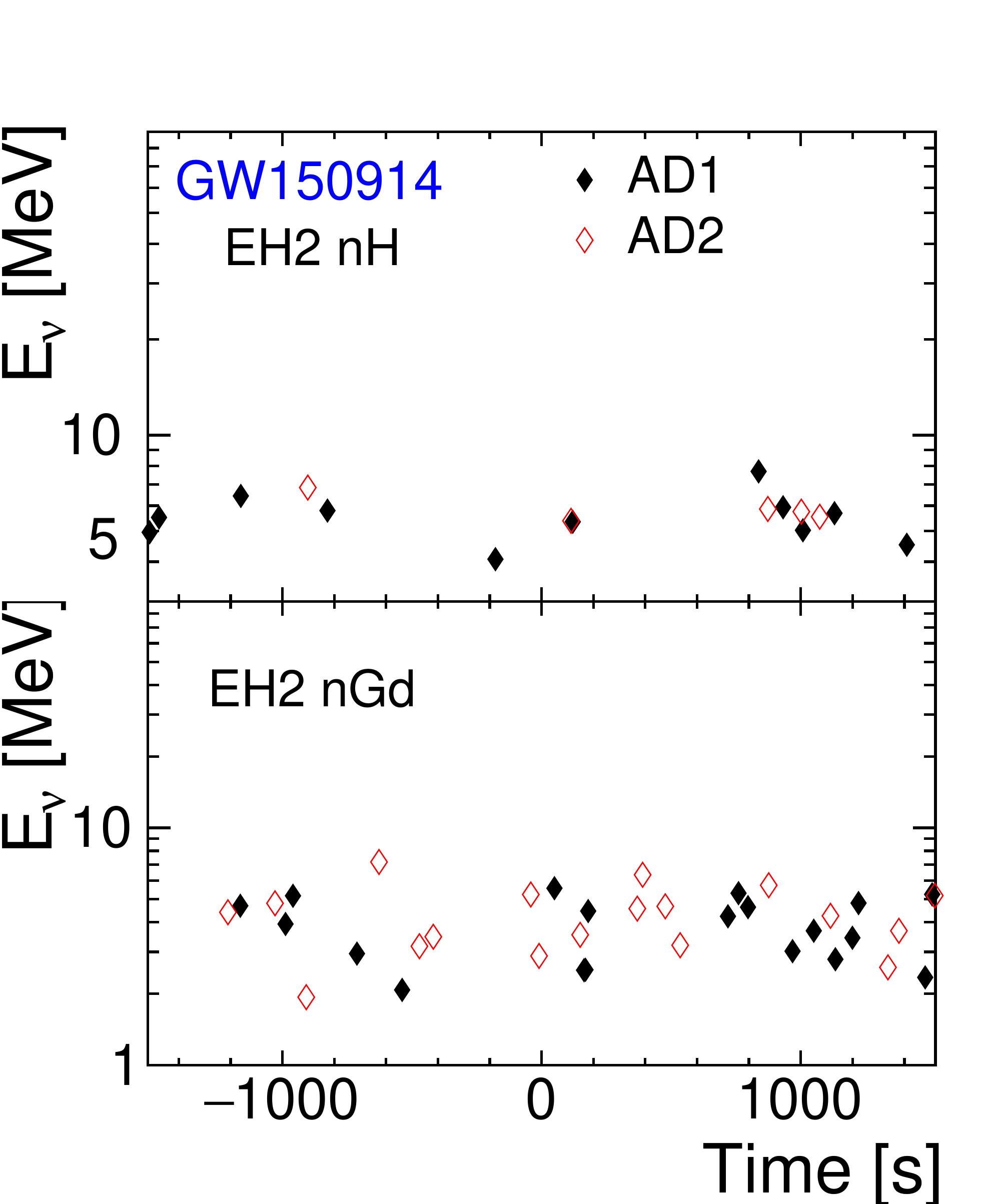}
\includegraphics[width=0.27 \textwidth]{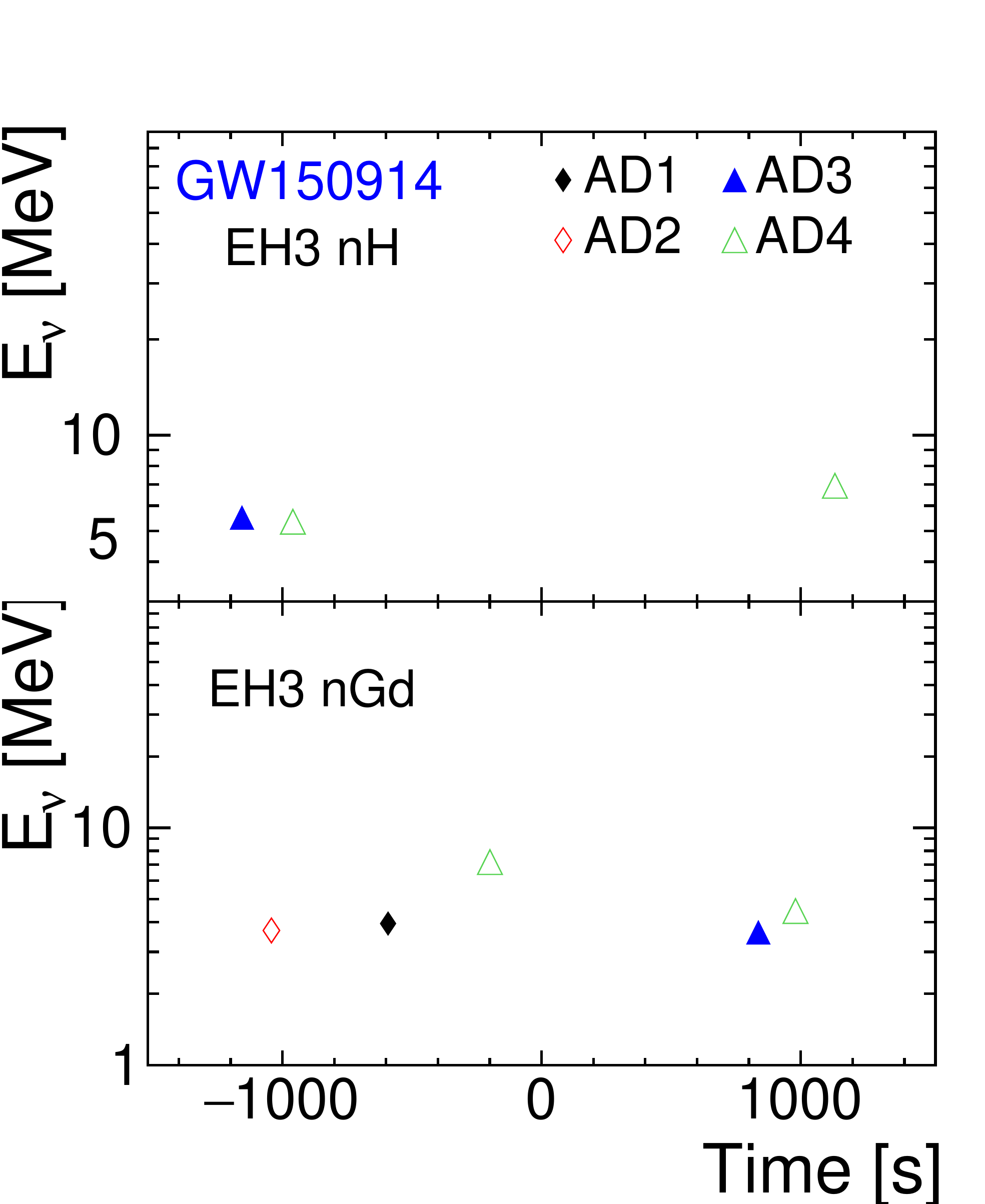}
\caption{Neutrino energy and relative time distribution of neutrino candidates for GW150914.}
\label{fig:Can1}
\end{figure}

\begin{figure}[htbp]
\centering
\includegraphics[width=0.27 \textwidth]{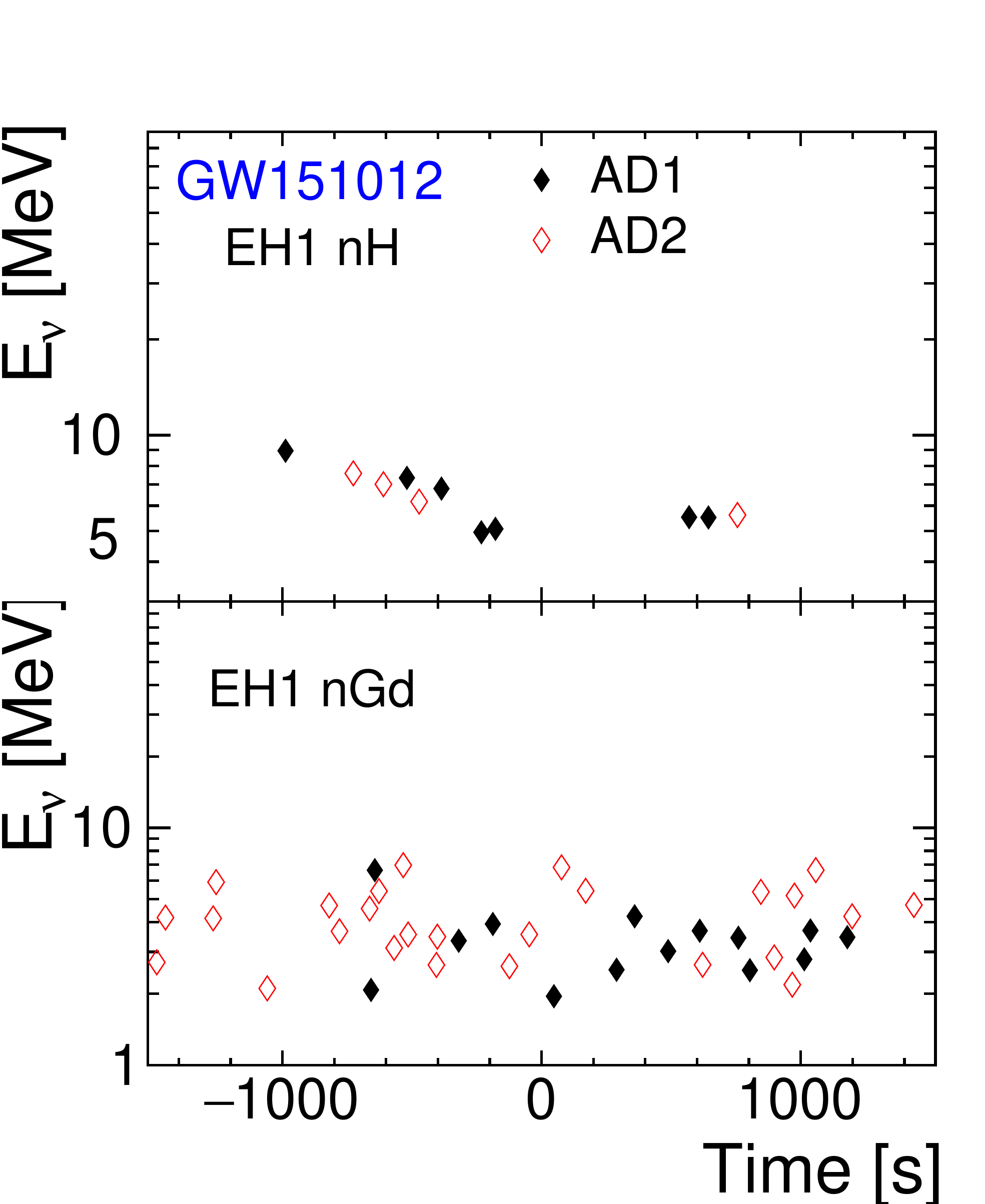}
\includegraphics[width=0.27 \textwidth]{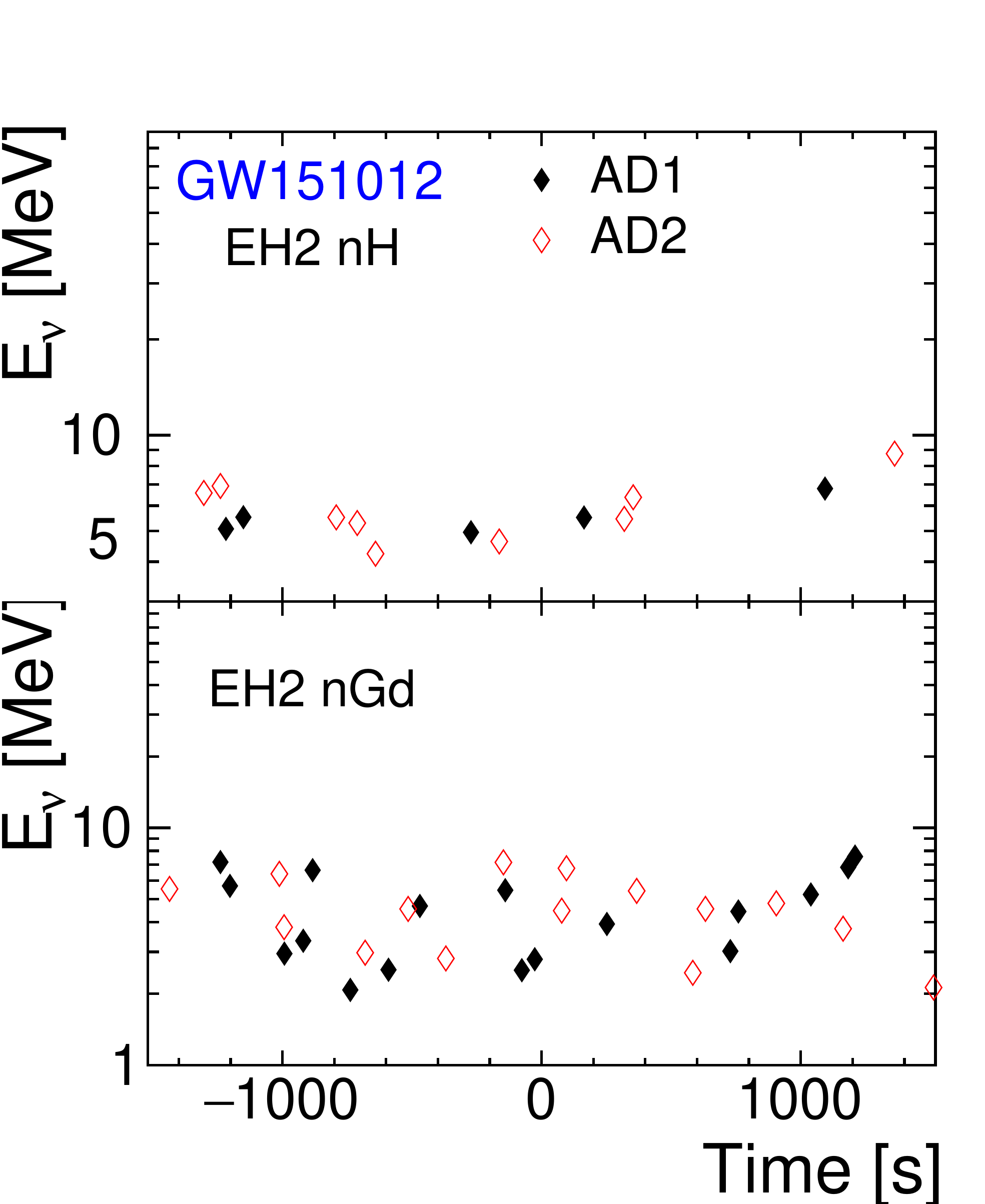}
\includegraphics[width=0.27 \textwidth]{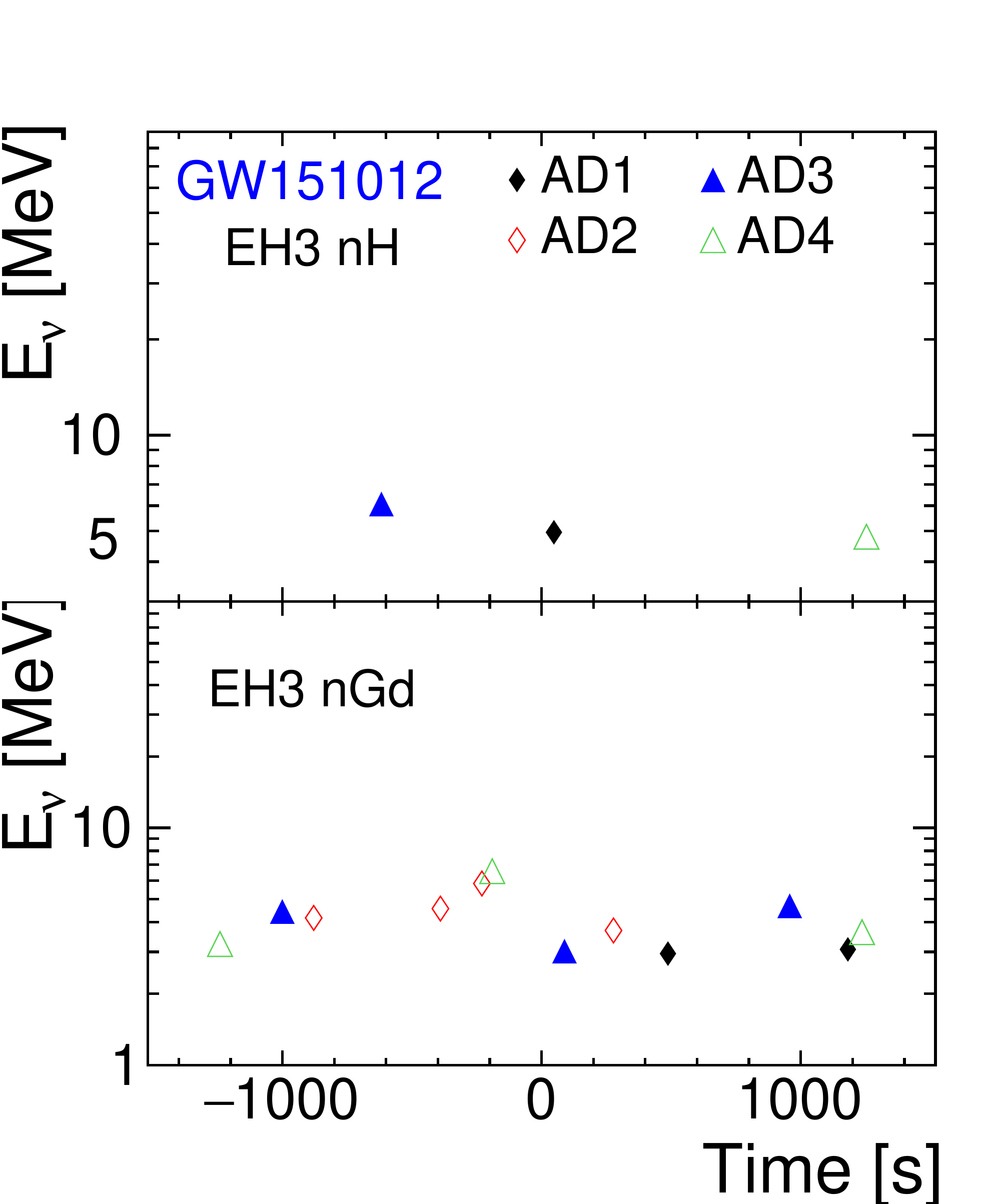}
\caption{Neutrino energy and relative time distribution of neutrino candidates for GW151012.}
\label{fig:Can2}
\end{figure}

\begin{figure}[htbp]
\centering
\includegraphics[width=0.27 \textwidth]{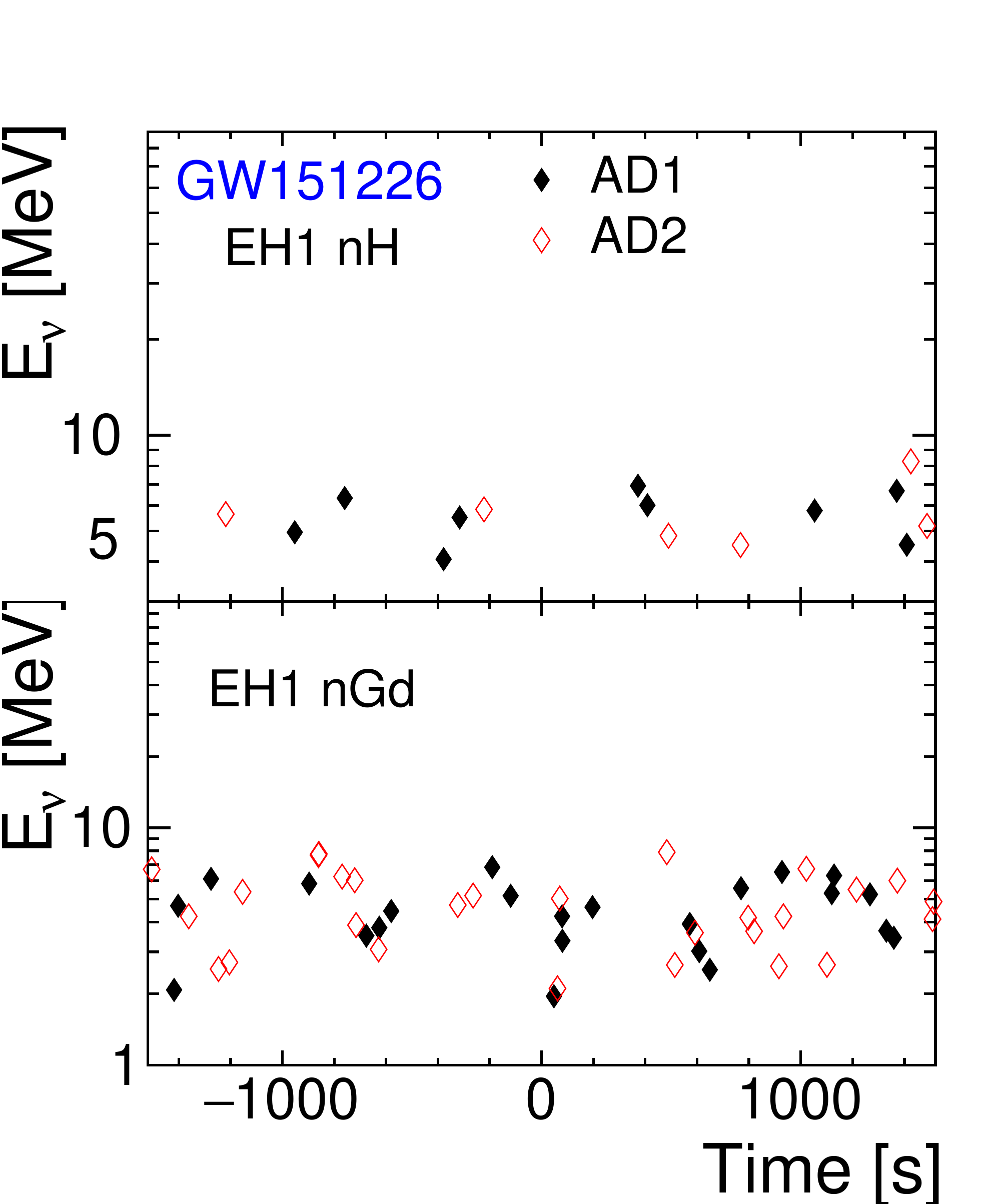}
\includegraphics[width=0.27 \textwidth]{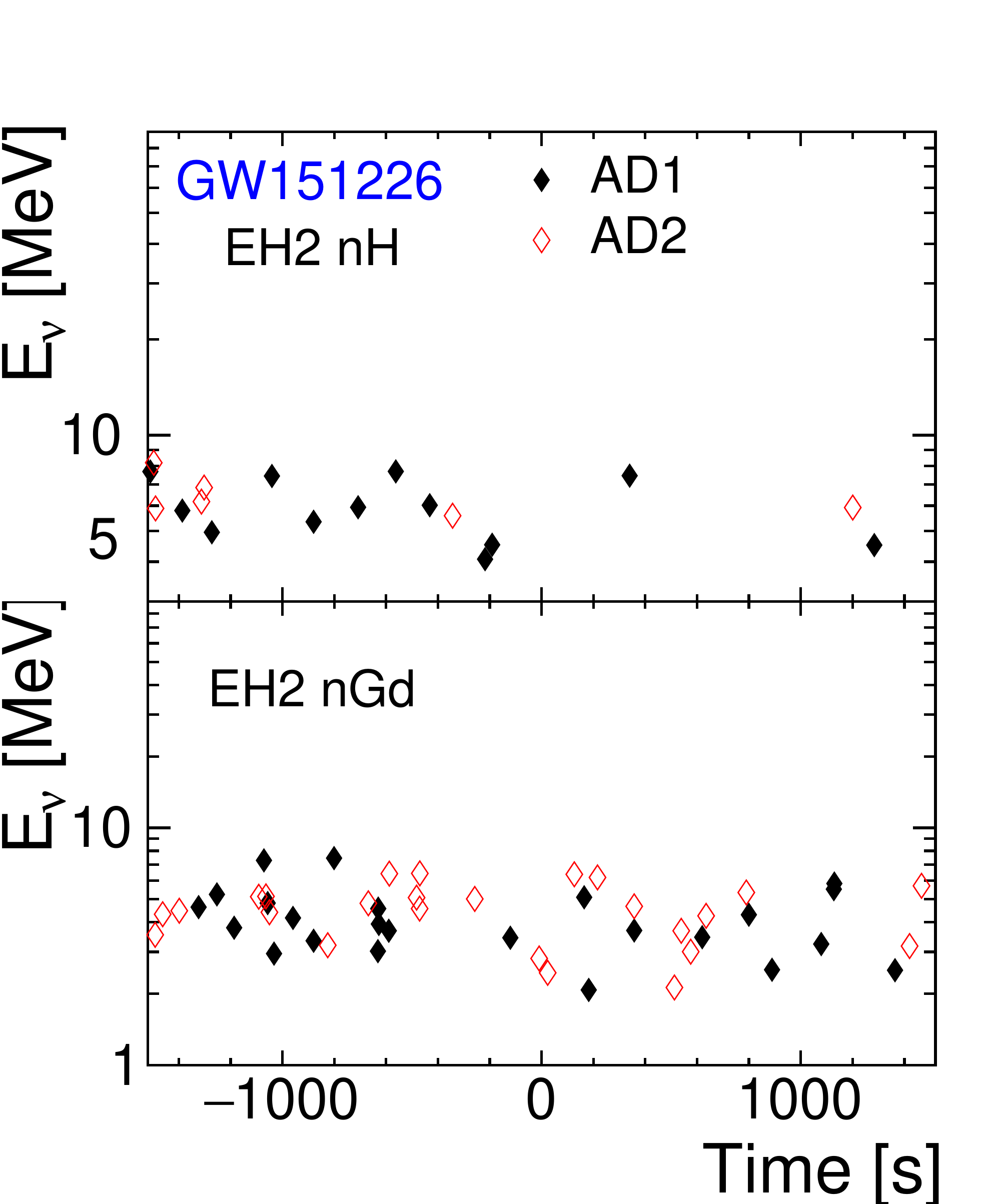}
\includegraphics[width=0.27 \textwidth]{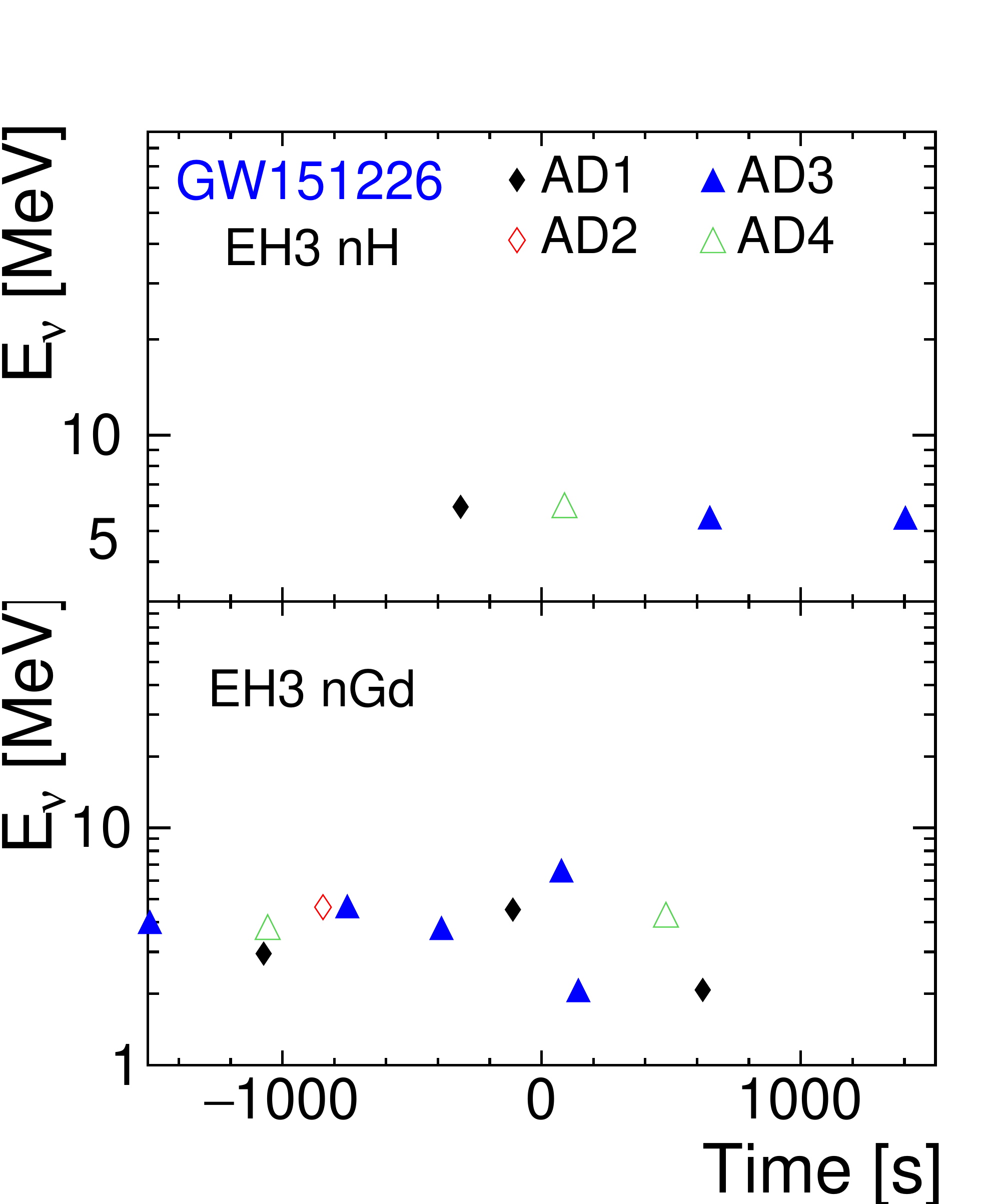}
\caption{Neutrino energy and relative time distribution of neutrino candidates for GW151226.}
\label{fig:Can3}
\end{figure}

\begin{figure}[htbp]
\centering
\includegraphics[width=0.27 \textwidth]{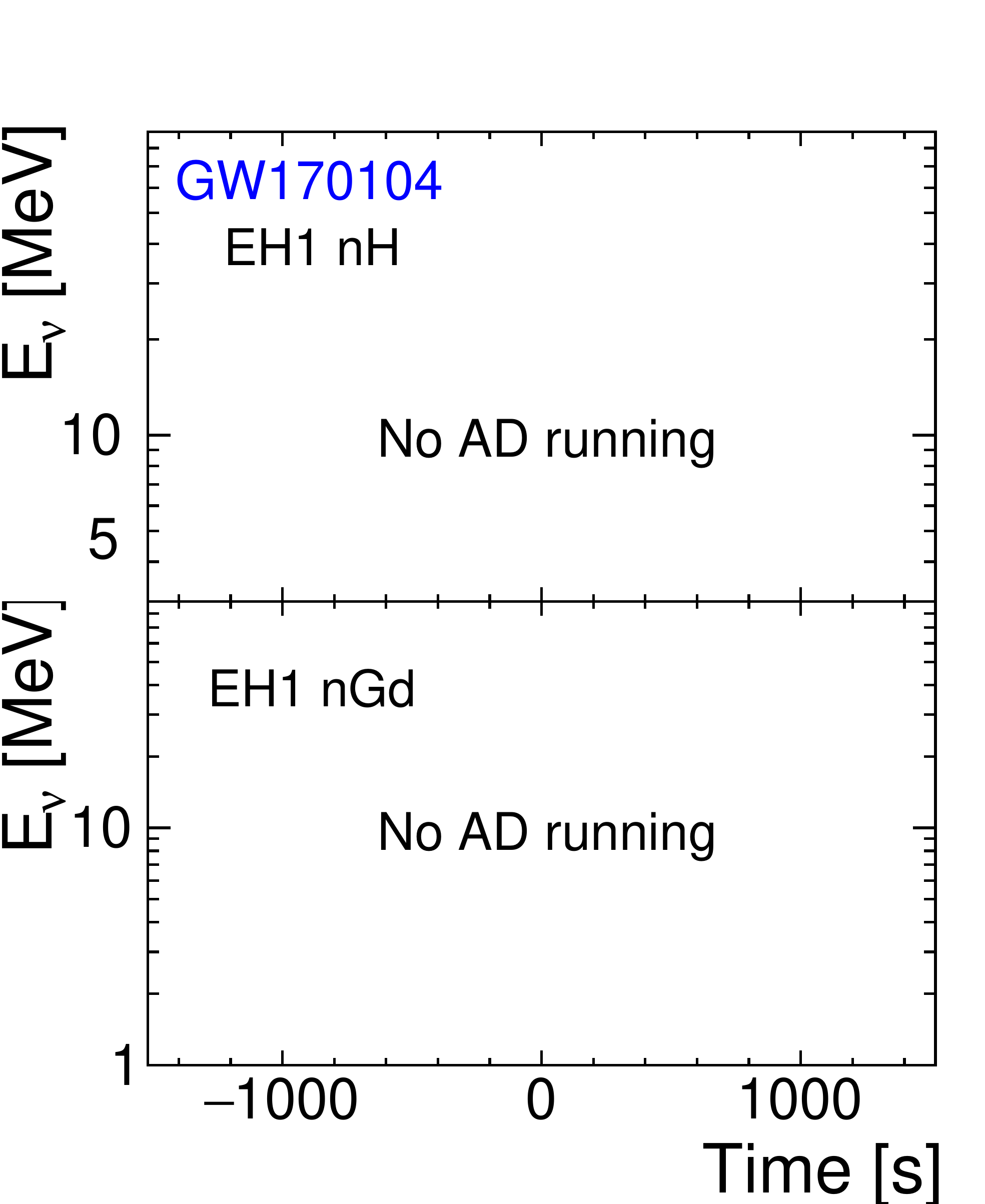}
\includegraphics[width=0.27 \textwidth]{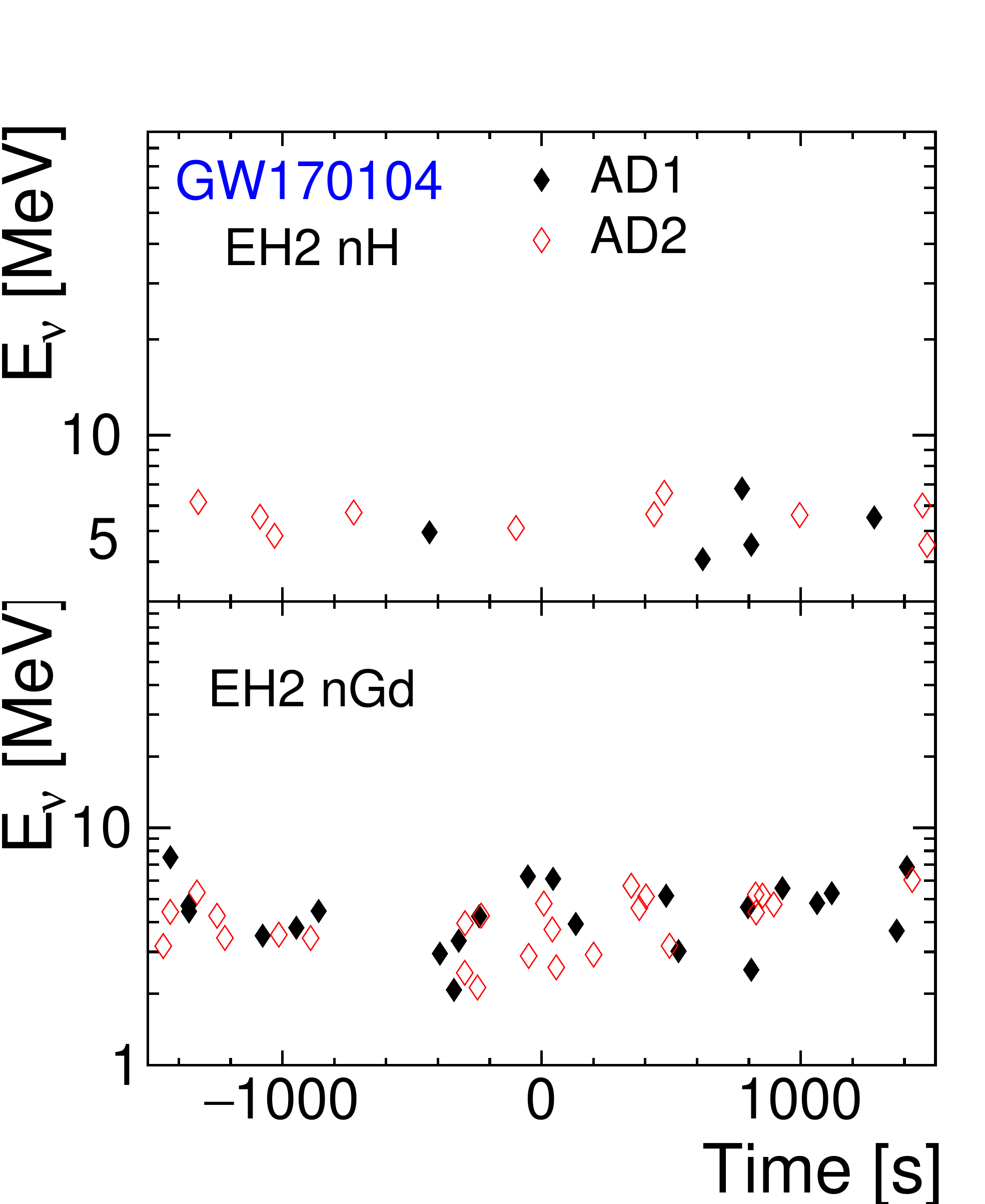}
\includegraphics[width=0.27 \textwidth]{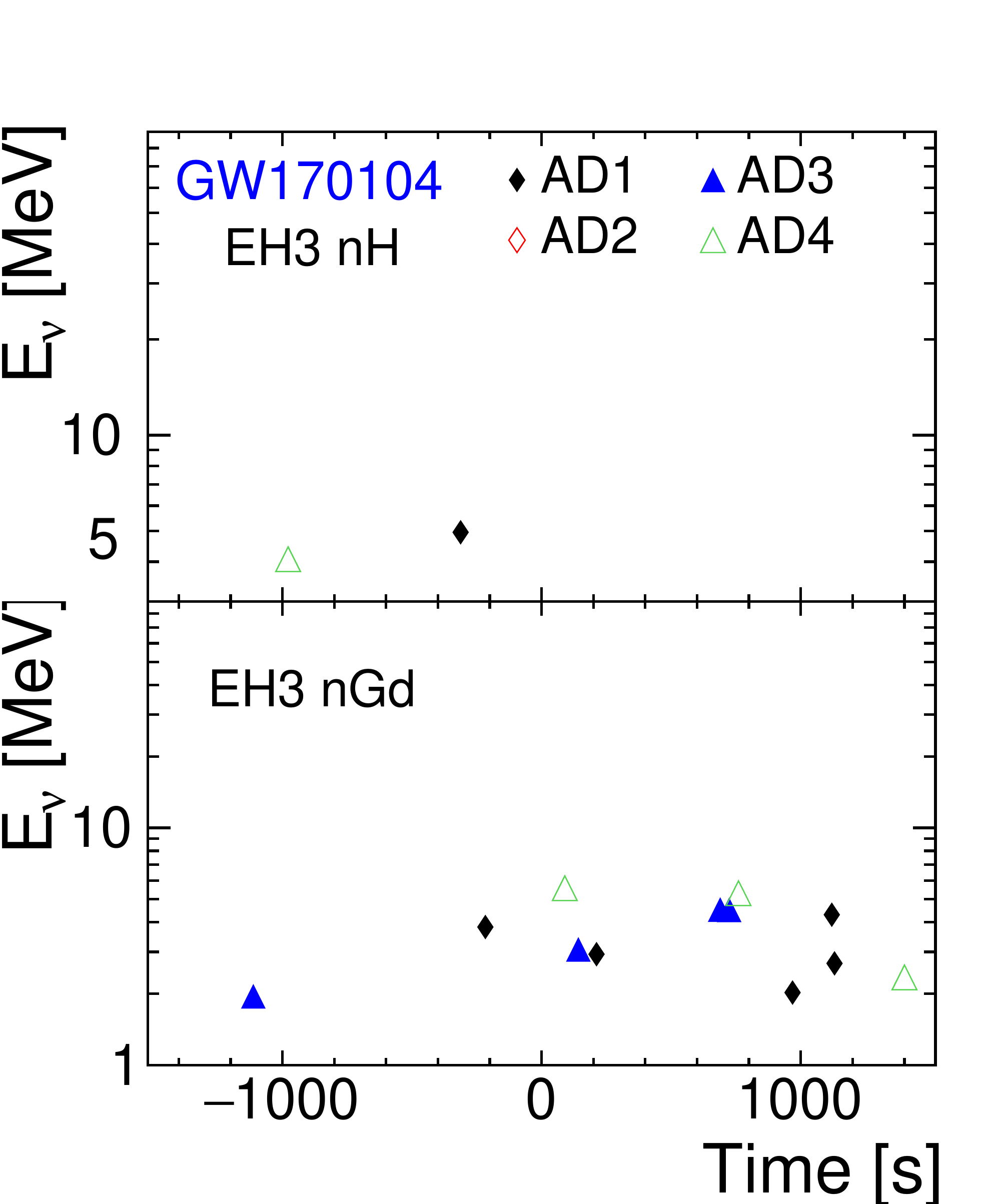}
\caption{Neutrino energy and relative time distribution of neutrino candidates for GW170104.}
\label{fig:Can4}
\end{figure}

\begin{figure}[htbp]
\centering
\includegraphics[width=0.27 \textwidth]{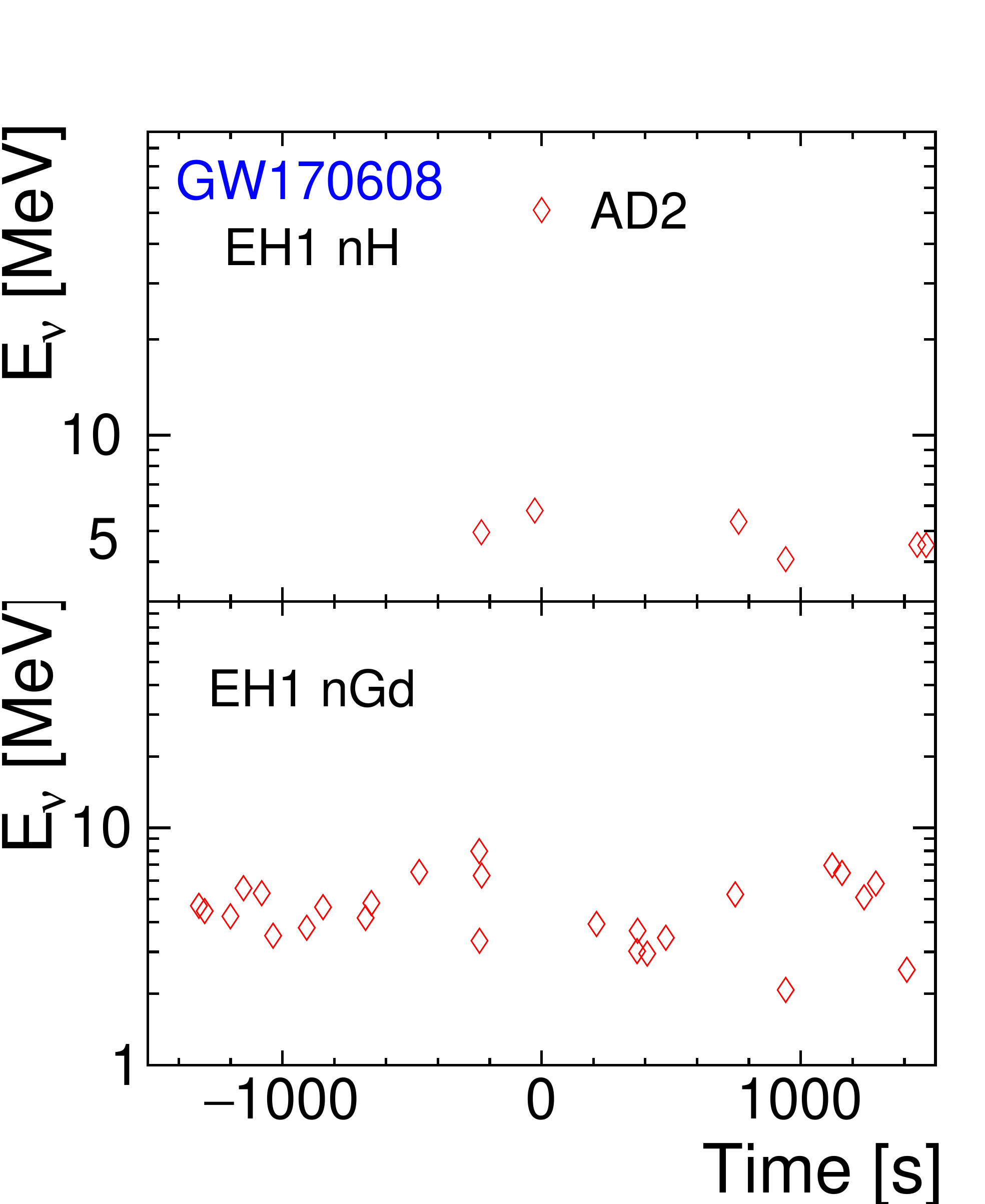}
\includegraphics[width=0.27 \textwidth]{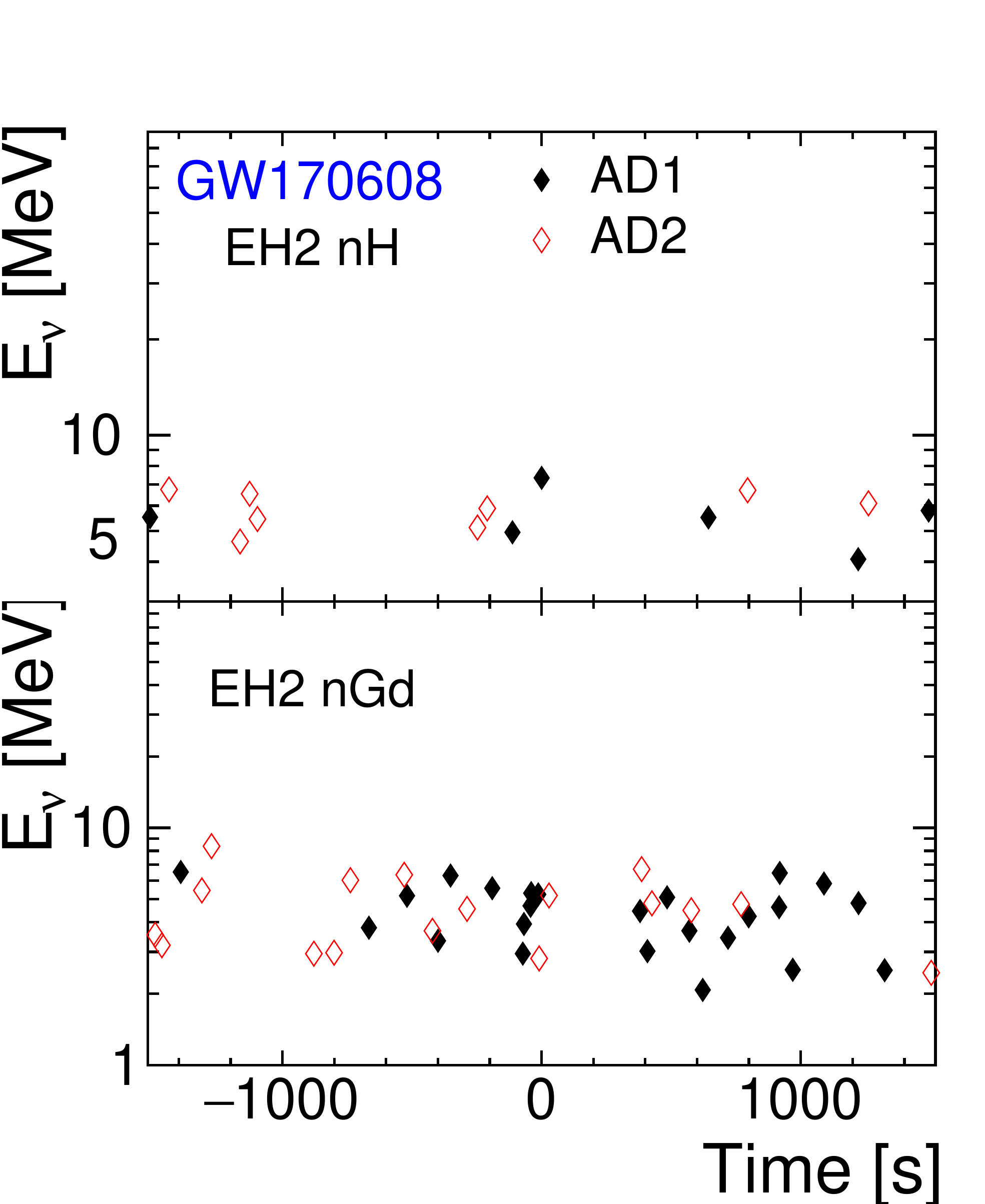}
\includegraphics[width=0.27 \textwidth]{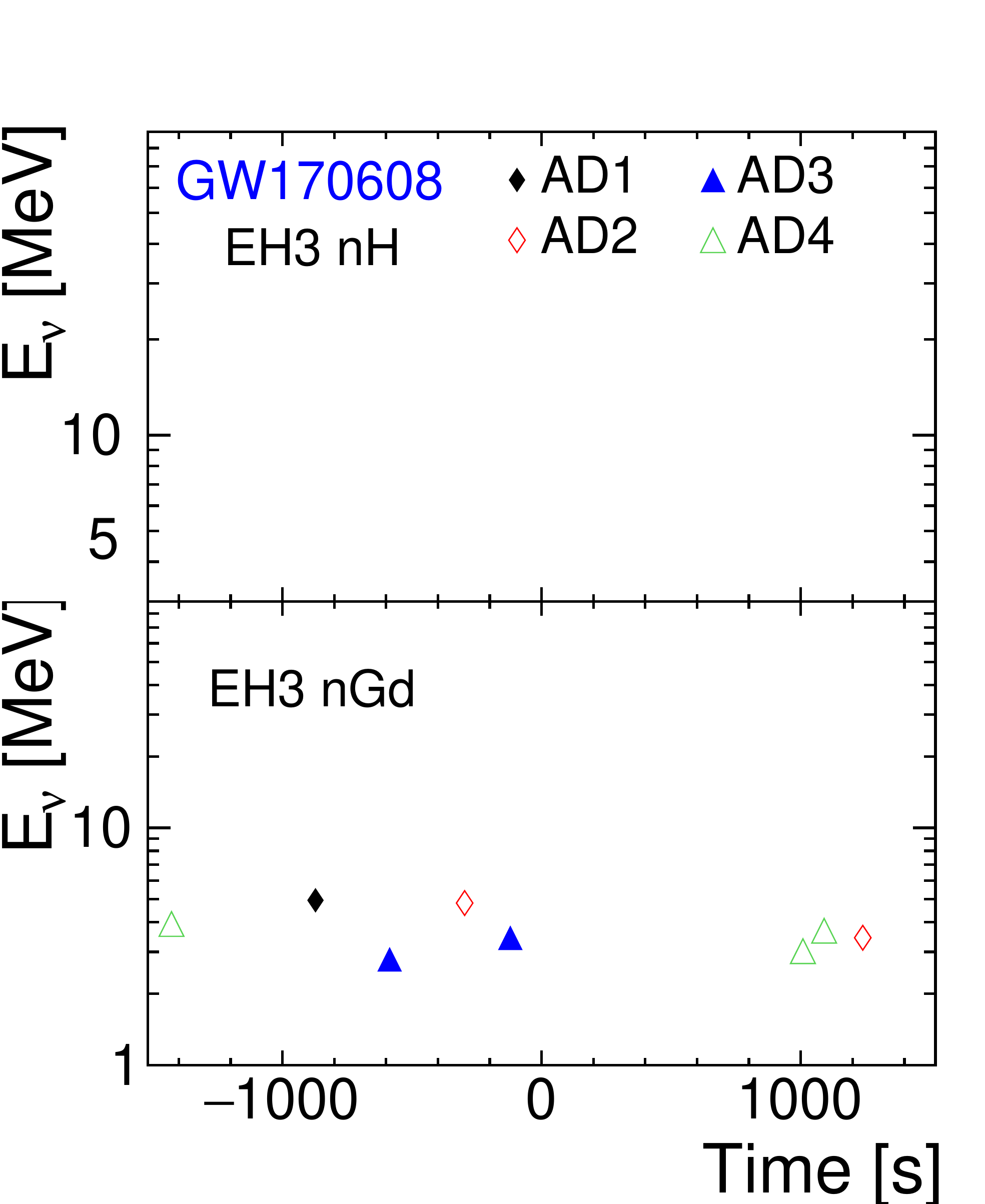}
\caption{Neutrino energy and relative time distribution of neutrino candidates for GW170608.}
\label{fig:Can5}
\end{figure}

\begin{figure}[htbp]
\centering
\includegraphics[width=0.27 \textwidth]{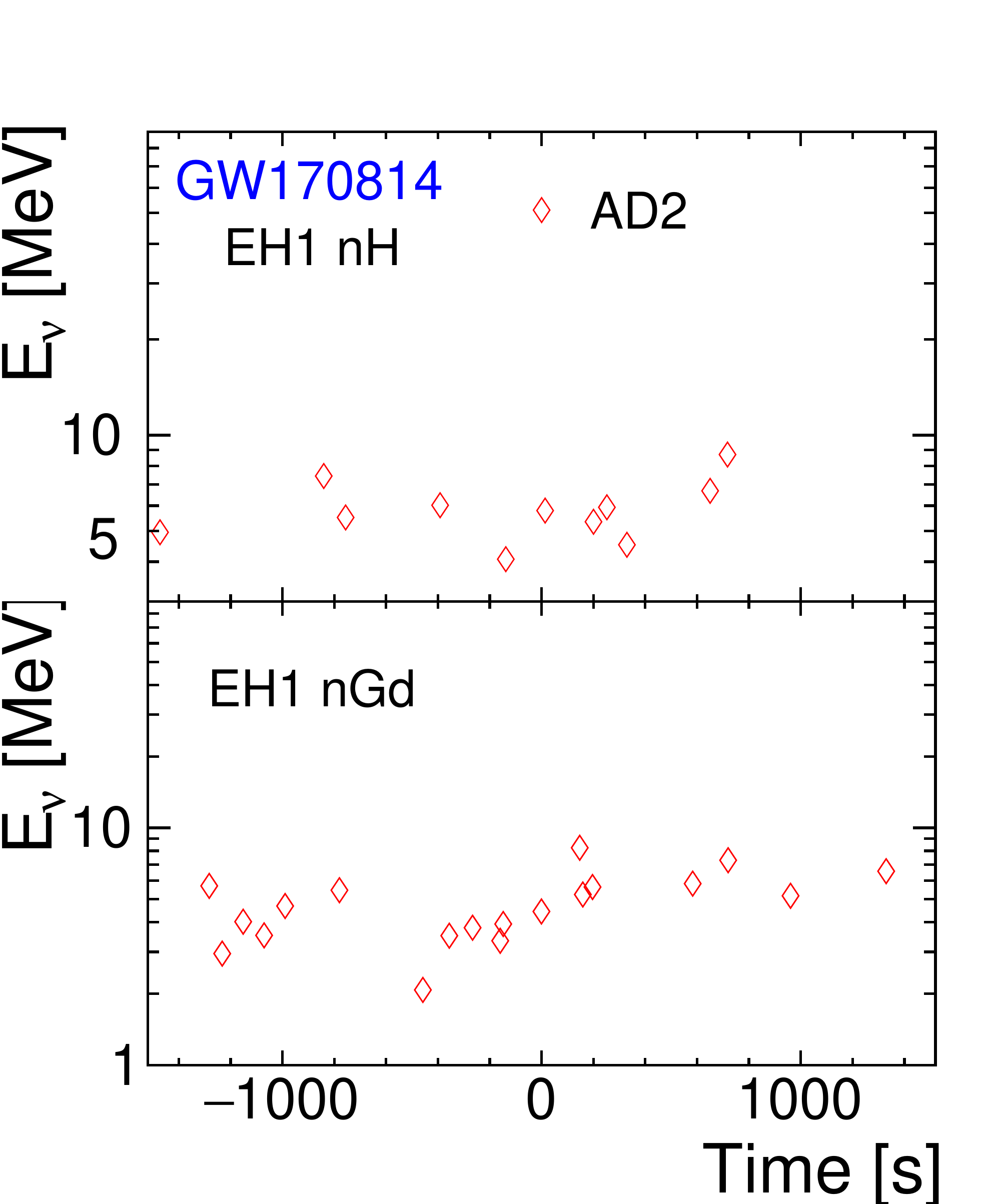}
\includegraphics[width=0.27 \textwidth]{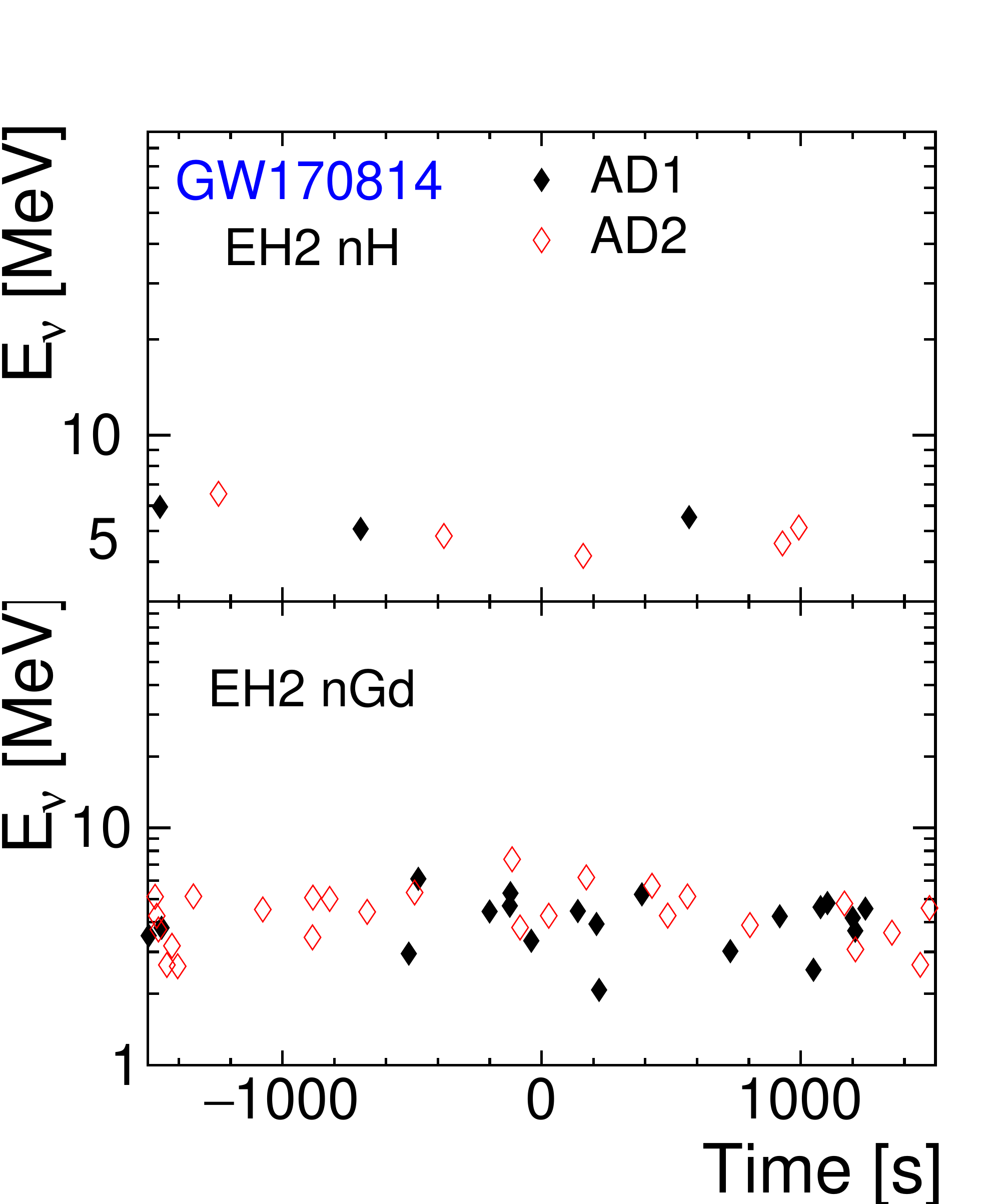}
\includegraphics[width=0.27 \textwidth]{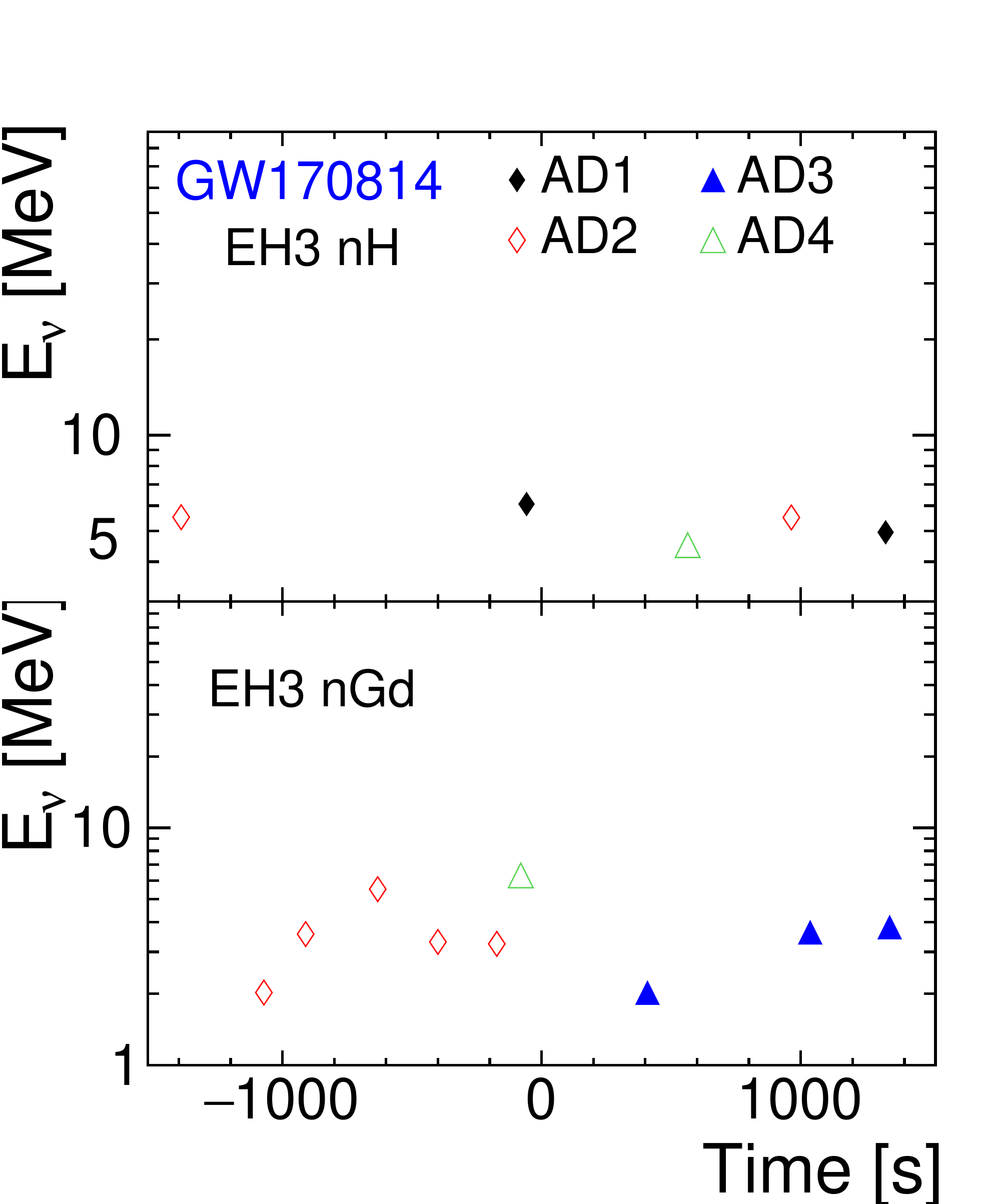}
\caption{Neutrino energy and relative time distribution of neutrino candidates for GW170814.}
\label{fig:Can6}
\end{figure}

\begin{figure}[htbp]
\centering
\includegraphics[width=0.27 \textwidth]{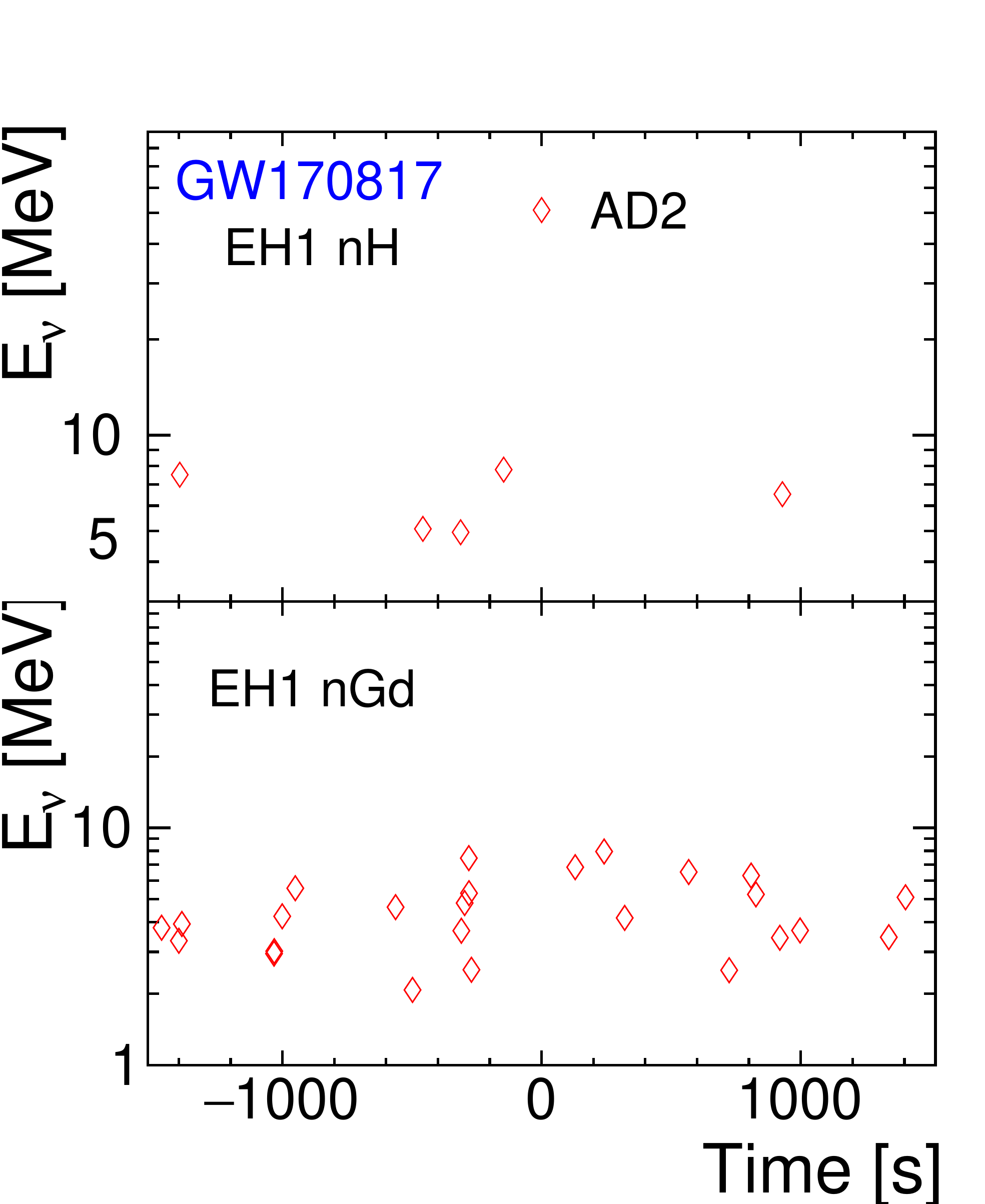}
\includegraphics[width=0.27 \textwidth]{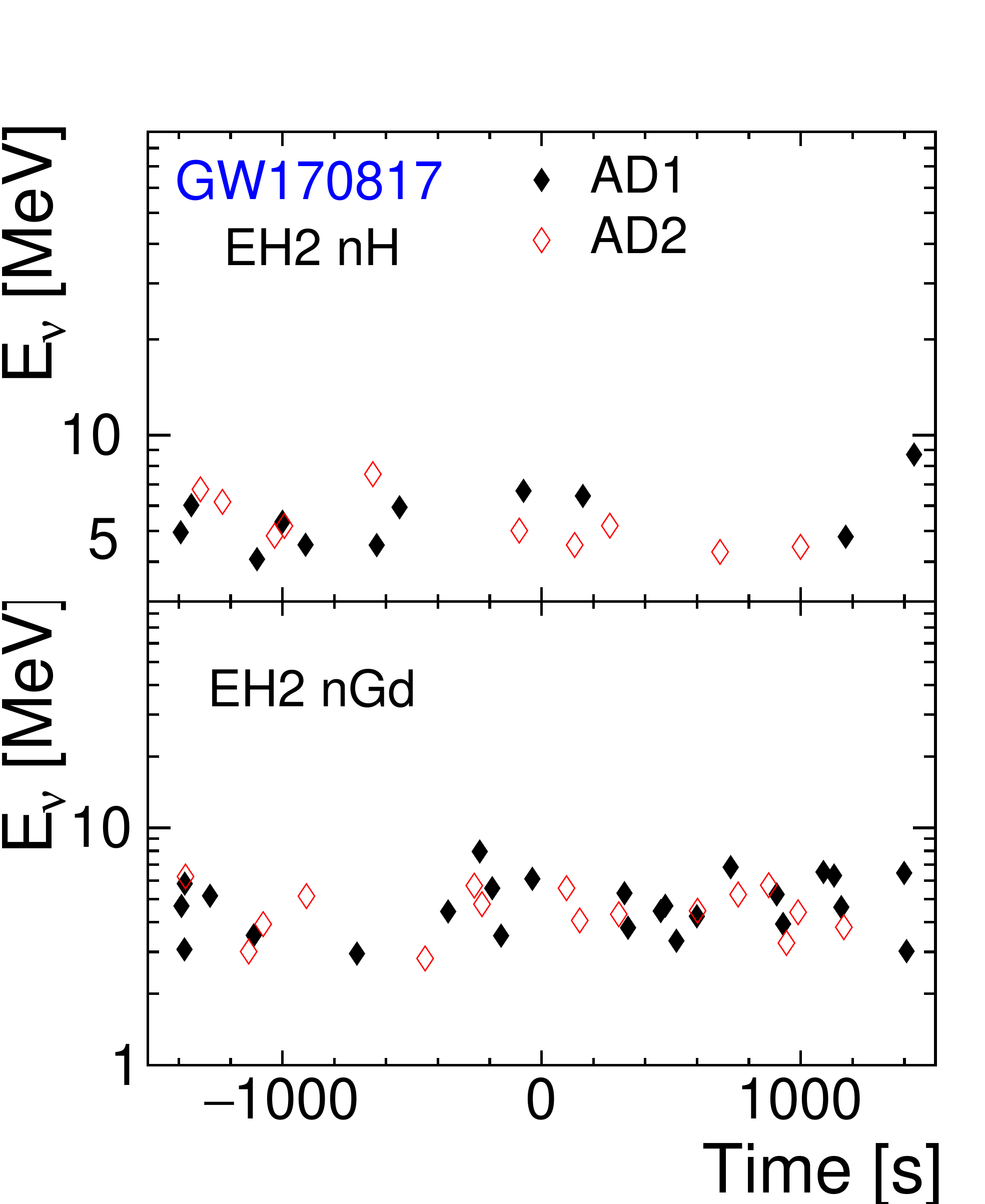}
\includegraphics[width=0.27 \textwidth]{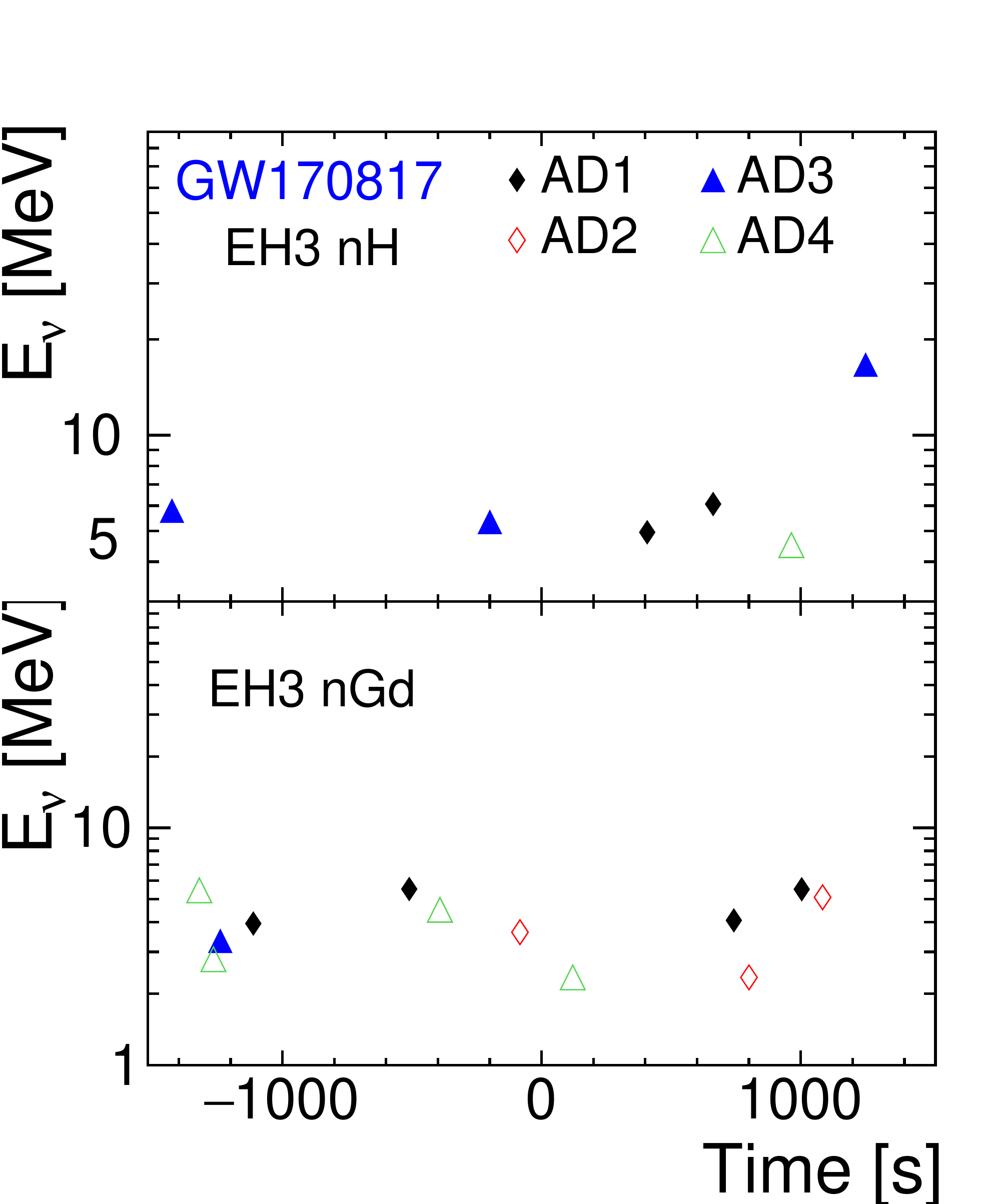}
\caption{Neutrino energy and relative time distribution of neutrino candidates for GW170817.}
\label{fig:Can7}
\end{figure}

Information regarding the selected neutrino candidates is shown in Figures~\ref{fig:Can1}, \ref{fig:Can2}, \ref{fig:Can3}, \ref{fig:Can4}, \ref{fig:Can5}, \ref{fig:Can6}, and \ref{fig:Can7}, where a two-dimensional plot of the measured neutrino energy $vs.$ relative time with respect to the GW observation time, is presented for each GW event and each experimental hall. Observations of $\pm 1500$ s are shown to see both the candidates ($\pm 10$ s, $\pm 500$ s, and $\pm 1000$ s) and backgrounds ([-1500 s, -1000 s] and [1000 s, 1500 s]) situation.




 \pagebreak

\bibliographystyle{apsrev4-1}
\bibliography{arXiv_GW_paper}

\end{document}